\documentclass[draftclsnofoot, onecolumn]{IEEEtran}
\usepackage{amsmath,amsfonts}
\usepackage{textcomp}

\usepackage{array,cite}
\usepackage[caption=false,font={small},labelfont={bf}]{subfig}
\usepackage{textcomp}
\usepackage{stfloats}
\usepackage{url}
\usepackage{enumitem}
\usepackage{verbatim}
\usepackage{cite}
\usepackage{balance}
\usepackage{enumitem}
\usepackage{booktabs}
\usepackage{multirow}
\usepackage{graphicx}
\usepackage{caption} 
\usepackage{amsthm}
\newtheorem{theorem}{Theorem}
\newtheorem{lemma}{Lemma}
\newtheorem{deft}{Definition}
\newtheorem{proposition}{Proposition}
\newtheorem{corollary}{Corollary}
\usepackage{xcolor}
\usepackage{tikz}
\usetikzlibrary{arrows.meta,calc,decorations.pathreplacing}

\newtheoremstyle{ieeeremark}
  {3pt}                    
  {3pt}                    
  {\normalfont}             
  {}                       
  {\bfseries}               
  {.}                       
  { }                       
  {}

\theoremstyle{ieeeremark}
\newtheorem*{Rem}{Remark}

\allowdisplaybreaks[4]

\usepackage{amsthm}
\usepackage{amssymb}

\usepackage{algorithm}
\usepackage{algpseudocode}  
\usepackage{hyperref}  
\theoremstyle{definition}
\newtheorem{example}{Example}

\begin{document}
\title{Moduli Selection in Robust Chinese Remainder Theorem: Closed-Form Solutions and Layered Design}
\author{Wenyi Yan, Lu Gan, Hongqing Liu, Shaoqing Hu, 
\thanks{This work has been submitted to the IEEE for possible publication. Copyright may be transferred without notice, after which this version may no longer be accessible.

Wenyi Yan, Lu Gan, and Shaoqing Hu are with the College of Engineering, Design and Physical Sciences, Brunel University of London, UK.  
Hongqing Liu is with the School of Communications and Information Engineering, Chongqing University of Posts and Telecommunications, China.  
Part of this work was presented at ICASSP 2024~\cite{yan2024towards}.}}

\maketitle

\begin{abstract}
We study the fundamental problem of \emph{moduli selection} in the Robust Chinese Remainder Theorem (RCRT), where each residue may be perturbed by a bounded error. Consider $L$ moduli of the form $m_i = \Gamma_i m$ ($1 \le i \le L$), where $\Gamma_i$ are pairwise coprime integers and $m \in \mathbb{R}^+$ is a common scaling factor. For small $L$ ($L = 2, 3, 4$), we obtain exact solutions that maximize the robustness margin under dynamic-range and modulus-bound constraints. We also introduce a Fibonacci-inspired \emph{layered} construction (for $L = 2$) that produces exactly $K$ robust decoding layers, enabling predictable trade-offs between error tolerance and dynamic range. We further analyze how robustness and range evolve across layers and provide a closed-form expression to estimate the success probability under common data and noise models. The results are promising for various applications, such as sub-Nyquist sampling, phase unwrapping, range estimation, modulo analog-to-digital converters (ADCs), and robust residue-number-system (RNS)-based accelerators for deep learning. Our framework thus establishes a general theory of moduli design for RCRT, complementing prior algorithmic work and underscoring the broad relevance of robust moduli design across diverse information-processing domains.
\end{abstract}


\section{Introduction}
The classical Chinese Remainder Theorem (CRT) states that an integer can be uniquely reconstructed from its residues modulo $L$ pairwise coprime integers $\{m_1,m_2,\ldots,m_L\}$, with the reconstruction guaranteed within the least common multiple (lcm) of the moduli. Owing to its algebraic structure, CRT underpins applications in digital arithmetic~\cite{381948}, cryptography~\cite{1190587}, coding theory~\cite{1190587, Mandelbaum76, BarsiMaestrini78, Shiozaki88, CRT_codes_TIT,Lattice_TIT}, sequence design for collision channels~\cite{IT_CRT_Seq,IT_collision_CRT} and secure communication~\cite{8664195}.

Despite its versatility, the classical CRT is inherently noise-sensitive. Within information theory, \emph{CRT codes} have been studied as error-correcting codes under an erasure error model, where some residues are received correctly while others may be erroneous~\cite{CRT_codes_TIT}. In practice, however, all residues may be corrupted, and robustness against such perturbations becomes essential. To address this challenge, Wang and Xia introduced Robust CRT (RCRT), which extends the classical model by allowing every residue to be perturbed by a bounded error~\cite{CRT}. Their framework guarantees a unique and stable reconstruction provided that the noise amplitude remains below a modulus-dependent threshold. Notably, RCRT can be formulated for both integer-valued and real-valued moduli and inputs, broadening its scope beyond discrete arithmetic and making it applicable to a wide range of problems, including sub-Nyquist sampling~\cite{xiao2017notes,1518897,RCRT4}, phase unwrapping~\cite{7230425}, range estimation~\cite{yang_phase_2014,akhlaq2016selecting,li2016wireless}, motion estimation~\cite{chi2024robust}, modulo analog-to-digital converters (ADCs)~\cite{gan_high_2020,gong_multi-channel_2021,yan2024towards,yan2025threshold}, and fault-tolerant analog neural network training~\cite{demirkiran2024blueprint,demirkiran2024mirage}.

Building on this foundation, numerous RCRT variants have been developed, including multi-stage grouping-based methods~\cite{xiao2014multi}, lattice-based approaches~\cite{Akhlaq2015BasisCF,akhlaq2016selecting,li2013distance,li2016wireless}, and search-based strategies~\cite{xiao_robustness_2018}. Extensions to multiple integer inputs~\cite{LiaoX07,LiaoX07}, multidimensional~\cite{xiao2020exact,xiao2024robust,guo2025construction} and complex-valued systems~\cite{gong_multi-channel_2021,li2025maximum} have also been explored. Among these, the layered RCRT model of Xiao \emph{et al.}~\cite{xiao_towards_2017} is particularly notable, as it introduces $K$ robust layers atop the classical CRT, enabling multi-level reconstruction that trades dynamic range for improved error tolerance. Together, these advances highlight the central importance of robustness in modern CRT-based systems.

Although substantial progress has been made on reconstruction algorithms, systematic moduli design remains an open challenge, with most existing approaches relying on heuristic parameter choices. Note that classical digital residue number systems (RNS) adopt structured modulus sets such as $\{2^n \!-\! 1, 2^n, 2^n+1\}$ to facilitate hardware efficiency~\cite{9459133,9177128,rns_review}. Although effective for error-free digital systems, such RNS constructions do not address robustness requirements in scenarios where \emph{all} residues are corrupted by noise. This limitation is particularly pronounced in emerging applications such as multi-channel modulo ADCs~\cite{gan_high_2020,gong_multi-channel_2021} and \emph{optical DNN accelerators}~\cite{demirkiran2024blueprint,demirkiran2024mirage}, which exploit CRT for parallelism and precision but remain vulnerable to rounding errors and hardware imperfections. These considerations motivate the development of moduli design strategies explicitly tailored for robustness.

This paper addresses the above-mentioned issues by developing explicit, closed-form moduli construction strategies tailored for RCRT systems in information processing. Our key contributions are:

\begin{itemize}
\item \textbf{Closed-form Moduli Design for the full CRT Layer:} We derive optimal constructions to select $L\in\{2, 3, 4\}$ moduli that maximize error tolerance at the full CRT layer. The solutions satisfy design constraints on modulus size and dynamic range with computational complexity $\mathcal{O}(1)$, enabling efficient hardware implementation in applications such as range estimation, modulo ADCs and analog neural network training.

\item \textbf{Layered Moduli Construction via Fibonacci-like Sequences:} We propose an explicit construction method for two-moduli systems supporting exactly $K$ robust layers. By leveraging Fibonacci-like integer sequences, the design guarantees controlled remainder chain length and facilitates predictable trade-offs between dynamic range and robustness.

\item \textbf{Analysis of Range–Robustness Trade-offs and Success Probability:} We analyze how the number of robust layers $K$ influences both dynamic range and error tolerance across layers. Moreover, we derive closed-form estimates for overall success probability under representative signal and noise models, providing valuable guidelines for system-level design.
\end{itemize}

The rest of this paper is organized as follows. Section~\ref{sec:fundamentals} reviews the principles of RCRT and multi-level reconstruction. Existing research on moduli selection is also discussed. Section~\ref{sec:opSe} presents closed-form moduli designs for the full CRT layer with \(L = 2,  3,\) and $4$ moduli. Section~\ref{sec:Multi-level} introduces the proposed Fibonacci-inspired construction for layered systems with two moduli. Section~\ref{sec:Layered} presents range–robustness trade-offs and derives probabilistic performance estimates. Section~\ref{sec:simul} validates the proposed methods through simulation and concludes in Section~\ref{sec:conclusion}.

\begin{table}[!t]
\renewcommand{\arraystretch}{1.3}
\caption{List of Notations}
\label{table:notation}
\centering
\begin{tabular}{l|p{5.6cm}}
\toprule
\hline
\textbf{Notation} & \textbf{Explanation} \\
\hline
$x$ / $\tilde{x}$ & Real-valued input signal / Reconstructed signal \\
$L$ & Number of moduli \\
$\Gamma_i$ & Coprime base moduli, for $1 < i \leq L$ \\
$m$ & Common scaling factor \\
$m_i$ & Moduli: $m_i = m\Gamma_i$  \\
$r_i$ / $\tilde{r}_i$ & Noiseless / noisy remainders modulo $m_i$ \\
$e_i$ & Additive error on $r_i$ \\
$n_i$ & True folding integers: $n_i = \lfloor x / m_i \rfloor$ \\
 $\tilde{n}_i$ & Estimated folding integers: $\tilde{n}_i = \lfloor \tilde{x} / m_i \rfloor$ \\
$N_{\mathrm{th}}$ & Target dynamic range threshold \\
$m_{\max}$ & Maximum allowed modulus \\
$\rho$ & Design ratio: $\rho = N_{\mathrm{th}} / m_{\max}$ \\
$\zeta_d$ & Multiplier in Fibonacci-based construction \\
$F_{d,k}$ & Fibonacci-like sequence with seed $(d, 1)$ \\
$K$ & Number of robust decoding layers \\
$\sigma_j$ & Remainder at layer $j$, for $1 \leq j \leq K+1$  \\
$P_j$ & Dynamic range at layer $j$ \\
$P_{K+1}$ & Final-layer dynamic range \\
$\tau_j$ & Tolerable error at layer $j$: $\tau_j = \frac{m\sigma_j}{4}$ \\
$T(x)$ & Piecewise error bound function \\
$\eta_{\mathrm{succ}}$ & Success probability of reconstruction \\
$\epsilon$, $\varsigma$ & Noise bound / standard deviation \\
$\varphi$ & Golden ratio: $\varphi = \frac{1 + \sqrt{5}}{2}$ \\
\hline
\bottomrule
\end{tabular}
\end{table}

\textbf{Notations:} The greatest common divisor and least common multiple of integers $a$ and $b$ are denoted by $\gcd(a, b)$ and $\text{lcm}(a, b) = |a b| / \gcd(a, b)$, respectively. Modulo operation is written as $ x \bmod m$, and the congruence relation as $a \equiv b \pmod{m}$. The floor, ceiling are denoted by $\lfloor \cdot \rfloor$ and $\lceil \cdot \rceil$. \(\mathbb{Z}_{> a}\) denotes the integers greater than \( a \); \(\mathbb{N}\), the positive integers. \(\mathcal{O}(\cdot)\) indicates computational complexity. The main variables and parameters used in this paper are summarized in Table~\ref{table:notation}.

\section{Literature Review}  
\label{sec:fundamentals}  
This section reviews the literature on RCRT systems, with particular emphasis on reconstruction methods that admit simple closed-form solutions, analogous to the classical CRT for integer inputs. Such methods are promising for real-time and low-latency applications.

Let \( x \geq 0 \) be a real-valued signal observed via \( L \) moduli \( \{m_i\}_{i=1}^L \). For each modulus \( m_i \), $x$ can be decomposed as \( x = n_i m_i + r_i \), where \( n_i = \lfloor x / m_i \rfloor \) is the folding integer and \( r_i = x \bmod m_i = x - n_i m_i \) is the remainder. In practical settings, however, the observed remainders are often corrupted by noise, yielding:
\begin{equation}
\tilde{r}_i = r_i + e_i = \left(x \bmod m_i \right)+ e_i,
\label{eq:errorresidue}
\end{equation}
where \( e_i \) is the residue (or remainder) error. Let \( \tilde{x} \) denote the reconstructed signal from the noisy remainders \( \{\tilde{r}_i\} \). In this context, we adopt the following definition of robust reconstruction~\cite{CRT}.

\begin{deft}[Robust Reconstruction]
Let \( \tilde{n}_i = \lfloor \tilde{x} / m_i \rfloor \) and \( n_i = \lfloor x / m_i \rfloor \) denote the estimated and true folding integers, respectively. Reconstruction is said to be robust if \( \tilde{n}_i = n_i \) for all \( i = 1,\ldots,L \).
\end{deft}

\subsection{Single-layer Standard RCRT systems}
 Wang~\emph{et al}. introduced a common factor \( m \) among the moduli~\cite{CRT}. Each modulus is written as \( m_i = m \Gamma_i \), where \( \Gamma_i \) are pairwise coprime integers. A simple, closed-form reconstruction was also developed in~\cite{CRT}. 
The following Proposition provides the theoretical guarantee for accurate reconstruction under bounded noise.

\begin{proposition}[Real-Valued RCRT~\cite{CRT}]\label{prop:classicalCRTbound}
Let the moduli be \( m_i = m\Gamma_i \) for \( 1 \le i \le L \), where \( m > 0 \) and \( \Gamma_1, \dots, \Gamma_L \) are pairwise coprime integers. Then for any \( x \in [0, P) \) with
\[
P = m \cdot \prod_{i=1}^L \Gamma_i,
\]
the folding integers \( n_i = \lfloor x / m_i \rfloor \) can be determined from noisy remainders \( \tilde{r}_i = x \bmod m_i + e_i \), provided the remainder errors satisfy
\begin{equation}\label{eq:diff-eq}
|e_1 - e_i| < \frac{m}{2}, \quad \text{for } 2 \le i \le L.
\end{equation}
A sufficient condition for \eqref{eq:diff-eq} is the uniform bound
\begin{equation} \label{eq:fullerror_Delta} 
|e_i| < \frac{m}{4}, \quad \text{for all } i,
\end{equation}
commonly referred to as the \textbf{error bound}.
\end{proposition}

To further improve robustness, Xiao \emph{et al.} developed a multi-stage RCRT framework~\cite{xiao2014multi} that relaxes the restrictive pairwise coprimality of \( \Gamma_i \). By grouping moduli into stages and applying RCRT hierarchically, this framework achieves larger error bounds and offers greater flexibility in system design.

Lattice-based CRT optimisation~\cite{Akhlaq2015BasisCF, akhlaq2016selecting, li2013distance, li2016wireless} reformulates the reconstruction as the closest vector search in a lattice defined by the moduli. These methods are applicable even when moduli are not co-prime and offer strong robustness under noise, but they incur significantly higher computational complexity compared to closed-form RCRT methods.

Similarly, Xiao \emph{et al.} proposed maximum likelihood estimation (MLE) based RCRT algorithms~\cite{xiao_robustness_2018,xiao2019solving}, which formulate reconstruction as a modular least-squares problem. These methods are highly flexible and robust to non-uniform and unbounded errors, but their exhaustive search nature leads to computational costs that scale poorly with dynamic range and the number of moduli, limiting practical deployment in real-time applications.

\subsection{Layered Robustness in RCRT Systems}

To address the trade-off between robustness and dynamic range, a~\emph{layered robustness} framework was developed for two-moduli systems~\cite{parhami_digital_2015, xiao_towards_2017}. The key idea is to partition the full dynamic range into multiple layers, each associated with a distinct error bound and dynamic range, as described below.

\begin{proposition}[Layered Dynamic Range for Two-Moduli RCRT~\cite{xiao_towards_2017}]
\label{prop:xiao_layered_DR}
Let the moduli be \( m_1 = m \Gamma_1 \) and \( m_2 = m \Gamma_2 \), where \( \Gamma_1 \) and \( \Gamma_2 \) are coprime positive integers and $m>0$ is the common scaling factor. Define \( \{\sigma_j\} \) recursively by
\begin{equation}\label{eq:sigmaexp_xiao}
   \sigma_{-1} = \Gamma_2,\ \sigma_0 = \Gamma_1,\ \sigma_j = \sigma_{j-2} \bmod \sigma_{j-1} \  (j \ge 1),
\end{equation}
and suppose this terminates at \( \sigma_{K+1} = 1 \).

Then, for any layer \( j \in \{1, \dots, K+1\} \), the folding integers \( n_1, n_2 \) can be uniquely recovered from noisy remainders \( \tilde{r}_i = x \bmod m_i + e_i \) if
\begin{equation} \label{eq:layered_error_Delta_xiao}
|e_1 - e_2| < \frac{m\sigma_j}{2}.
\end{equation}
The corresponding dynamic range at layer \( j \) is
\begin{equation}\label{eq:pj_xiao_expr}
P_j = m \cdot \min\left( \Gamma_2 (1 + \ddot{n}_{2,j}),\; \Gamma_1 (1 + \ddot{n}_{1,j}) \right),
\end{equation}
where \( \ddot{n}_{i,j} \) are the maximum correctly decodable folding integers at layer \( j \), given in~\cite[Eqs.~(51), (52)]{xiao_towards_2017}.
\end{proposition}

Condition~\eqref{eq:layered_error_Delta_xiao} holds whenever \( |e_i| < \frac{m}{4} \sigma_j \) for \( i = 1, 2, \) which is known as the \textit{layer-\(j\)} error bound. An alternative expression for \( P_j = m N_j \) was provided in~\cite{xiao_robustness_2018}, where
\begin{equation}\label{eq:Nj_xiao_alt}
\left\{
\begin{aligned}
&N_{-1} = \Gamma_1,\quad N_0 = \Gamma_2,\\
&N_j = N_{j-2} + \left\lfloor \frac{\sigma_{j-1}}{\sigma_j} \right\rfloor (N_{j-1} - \sigma_j), \quad j \ge 1.
\end{aligned}
\right.
\end{equation}

This layered RCRT architecture enables a controlled trade-off between error tolerance and dynamic range. In addition to the full CRT range \( P = m \Gamma_1 \Gamma_2 \), it provides \( K \) intermediate robust layers, each capable of tolerating larger errors at the cost of a reduced dynamic range. Such a multi-resolution representation is particularly beneficial in noisy environments such as ADCs and range estimation.

\begin{figure}[!t]
\centering
\hspace{-1cm}
\subfloat[]{\includegraphics[width=0.32\linewidth]{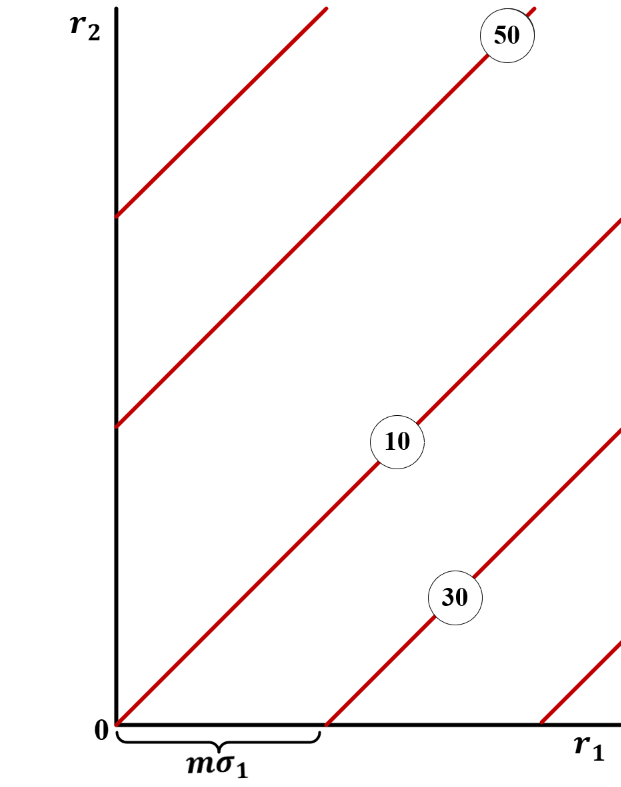}%
\label{fig:P1}} 
\hfill
\subfloat[]{\includegraphics[width=0.32\linewidth]{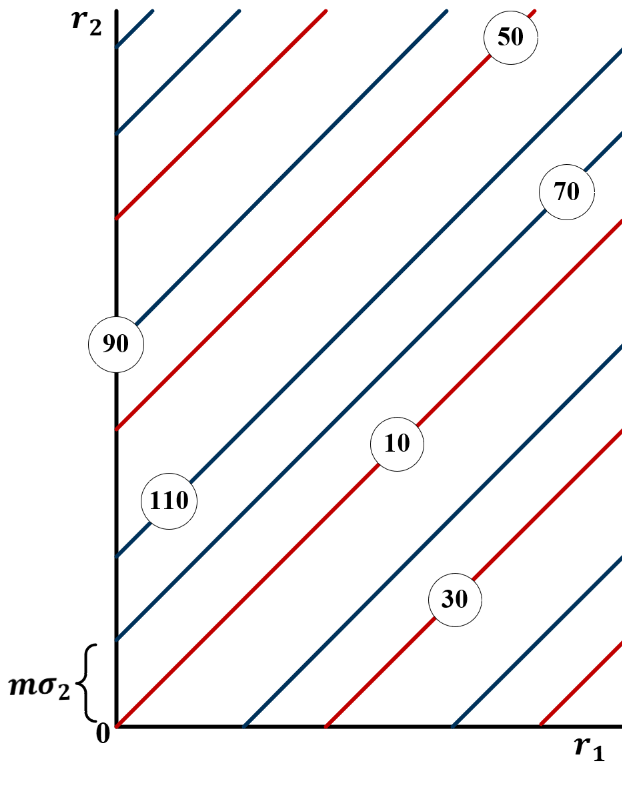}%
\label{fig:P2}}
\hfill
\subfloat[]{\includegraphics[width=0.32\linewidth]{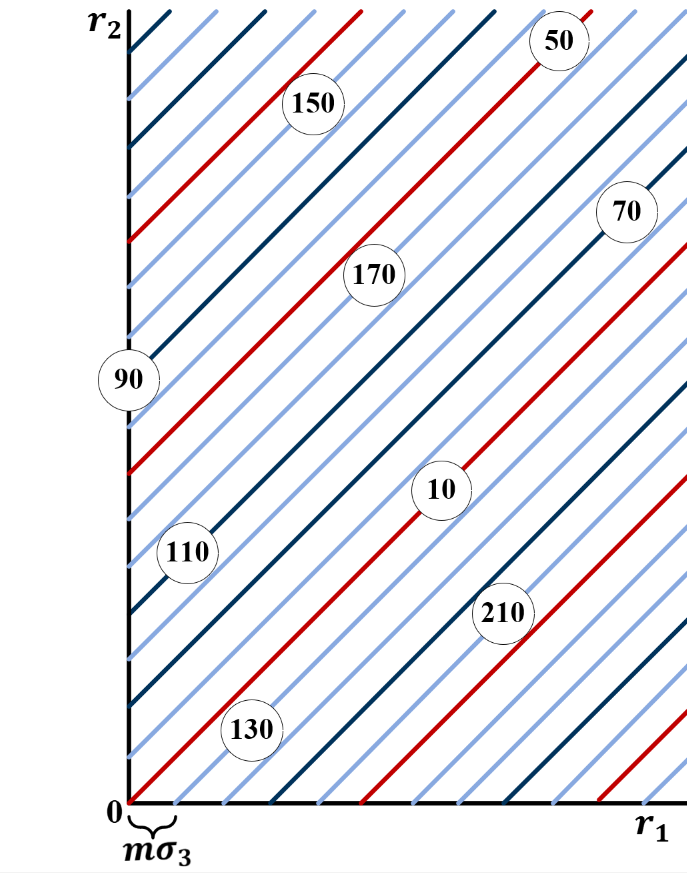}%
\label{fig:P3}}
\caption{Layered RCRT residue maps in the $(r_2,r_1)$ plane for $\Gamma_1=12$, $\Gamma_2=17$, $m=1.5$, and $K=2$.
Each plotted point is $(r_2,r_1)$ generated by a value $x$, with
$r_\ell = x \bmod m_\ell$ and $m_\ell=\Gamma_\ell m$ ($\ell=1,2$).
Small circled labels indicate the corresponding $x$.
Slanted families depict the pairwise-difference structure; at layer $j$ the inter-family spacing equals $m\sigma_j$ and the admissible dynamic range is $[0,P_j)$.
(a) Robust layer~1: $x\in[0,54]$, $m\sigma_1=7.5$.
(b) Robust layer~2: $x\in[0,127.5]$, $m\sigma_2=3$.
(c) Full CRT layer: $x\in[0,306]$, $m\sigma_3=1.5$.}
\label{fig:LayeredRCRT_example}
\end{figure}

\noindent\textbf{Geometric intuition (pairwise–difference view).}
Eq.~\eqref{eq:layered_error_Delta_xiao} admits a minimum–separation interpretation in the two–moduli residue plane. 
Fig.~\ref{fig:LayeredRCRT_example} illustrates $(r_1,r_2)$ for $m=1.5$, $\Gamma_1=12$, $\Gamma_2=17$.
The Euclidean remainders are $\sigma_1=5$, $\sigma_2=2$, and $\sigma_3=1$, i.e., $K=2$ robust layers followed by a full-CRT layer.
Each valid pair of folding integers $(n_1,n_2)$ generates a family of slanted lines
$
r_2 \;=\; r_1 + mc,\qquad c\in\mathbb{Z},
$
with \emph{non-uniform} gaps whose \emph{local} minimum separation at layer $j$ is lower-bounded by $m\,\sigma_j$.
Consequently, the decoder succeeds whenever the pairwise residue perturbation satisfies
~\eqref{eq:layered_error_Delta_xiao}.
Fig.~\ref{fig:LayeredRCRT_example} (a)–(c) visualize how the admissible dynamic range expands from $[0,54)$ to $[0,127.5)$ and finally to $[0,306)$, while the guaranteed local separation (and thus the error tolerance) tightens from $m\,\sigma_1=7.5$ to $m\,\sigma_2=3$ and then to $m\,\sigma_3=1.5$.

\subsection{Moduli Selection in RCRT Systems}
Moduli selection plays a critical role in both performance and hardware efficiency for RCRT-based systems. Classical integer CRT and RNS often adopt structured modulus sets near powers of two to enable fast arithmetic and binary-friendly implementations~\cite{rns_review}. While effective for digital computation, these designs do not account for robustness against noise or perturbations, and their coarse granularity limits applicability in analog domains such as modulo ADCs~\cite{yan2024towards}.

For  RCRT with a single full CRT layer, prior work has explored moduli selection for various applications. Xiao \emph{et al.}~\cite{xiao_robustness_2018} studied random prime-based selection to achieve statistical robustness in residue coding. Akhlaq \emph{et al.}~\cite{Akhlaq2015BasisCF,akhlaq2016selecting} proposed search-based wavelength selection schemes for least square-based range estimation by optimizing the underlying lattice geometry. However, these methods rely on solving the Closest Vector Problem (CVP) in high-dimensional lattices, which is computationally expensive and lacks deterministic error bounds. Moreover, optimized moduli often require high-precision parameters (e.g., Examples D in~\cite{akhlaq2016selecting}), which pose challenges for hardware implementation. Even for single-layer systems, none of these approaches yields explicit, closed-form strategies for moduli design within the RCRT framework.

To our knowledge, no existing work addresses moduli selection for \textit{layered RCRT systems}, where additional robust layers are introduced to improve error tolerance at the cost of reduced dynamic range. This gap highlights the need for closed-form, analytically grounded moduli design techniques applicable to both single-layer and layered RCRT. Developing such methods is essential for enabling efficient reconstruction and hardware implementation, which is the core motivation of this work.

\section{Moduli selection for Multi-modular Systems}
\label{sec:opSe}
In applications of RCRT systems (e.g., modulo ADCs, sub-Nyquist sampling, phase unwrapping, analog neural network training), moduli selection must satisfy:  
(\emph{i}) a hardware-imposed maximum modulus~\( m_i \leq m_{\max} \) and  
(\emph{ii}) an application-driven dynamic range threshold~\( P \geq N_{\text{th}} \).  

To enable efficient and closed-form reconstruction algorithms (e.g., based on RCRT~\cite{CRT}), we also require $\Gamma_i$ ($1\le i \le L$) to be \emph{pairwise coprime}. Although some prior studies (e.g.,~\cite{xiao2014multi,xiao_towards_2017,akhlaq2016selecting}) have explored non-coprime $\Gamma_i$ with more advanced decoding, such approaches typically involve complicated residue grouping or solving CVP problems in lattice search, making them less attractive for real-time applications.

Given a maximum modulus \( m_{\max} \) and a dynamic range threshold \( N_{\text{th}} \), the goal is to select \( L \) strictly increasing integers \( \Gamma_1 < \Gamma_2 < \cdots < \Gamma_L \), and define the moduli as
\begin{equation}
m_i = m \cdot \Gamma_i, \quad i = 1, \dots, L,
\end{equation}
subject to the following constraints:
\begin{align}
\textbf{Pairwise coprimality:} &\quad \gcd(\Gamma_i, \Gamma_j) = 1 \quad \forall \, i \neq j, \label{eq:pairwise_coprime}\\
\textbf{Modulus threshold:} &\quad m \cdot \Gamma_L \leq m_{\max}, \label{eq:threshold} \\
\textbf{Dynamic range:} &\quad m \cdot \prod_{i=1}^L \Gamma_i \geq N_{\text{th}}. \label{eq:range_constraint}
\end{align}

By Proposition~\ref{prop:classicalCRTbound}, the error bound is proportional to the common scaling factor \( m \), motivating its maximisation for noise robustness. From~\eqref{eq:threshold}, we set  
\begin{equation}
m = \frac{m_{\max}}{\Gamma_L}. \label{eq:mval}
\end{equation}
Substituting into~\eqref{eq:range_constraint} yields 
\begin{equation}
\prod_{i=1}^{L-1} \Gamma_i \geq \rho, 
\quad \text{where} \quad 
\rho = \frac{N_{\text{th}}}{m_{\max}}. \label{eq:rho_def}
\end{equation}
Since \( m \propto 1/\Gamma_L \), the design reduces to
\begin{equation}
\min_{2\le\Gamma_1 < \cdots < \Gamma_L} \quad \Gamma_L,
\label{eq:opt_problem}
\end{equation}
subject to the \emph{pairwise coprimality} condition~\eqref{eq:pairwise_coprime} and the \emph{product} constraint~\eqref{eq:rho_def}.

Finding an optimal solution to the above problem for an arbitrary $L$ is computationally intractable due to the combinatorial nature of the pairwise coprimality and product constraints. However, in many practical RCRT applications, the number of moduli $L$ is typically small due to hardware limitations. For instance, in multi-channel modulo ADCs~\cite{gan_high_2020}, each channel samples at the Nyquist rate $f_{\text{NYQ}}$, yielding a total sampling rate of $L f_{\text{NYQ}}$. The analog modulo operation~\cite{bhandari2021unlimited,mulleti2023hardware} on each channel requires digital counters and comparators, with complexity growing quickly as $L$ increases. On platforms constrained by size, weight, or power, such as implantable devices or portable radio frequency sensors, the number of parallel channels is therefore limited to small $L$. Similar restrictions arise in RCRT-based frequency and phase estimation~\cite{yang_phase_2014,chi2024robust}, where each modulus requires an additional sensing or measurement channel, making large $L$ impractical in radar, sonar, or communication systems. An analogous limitation is present in emerging analog neural network training accelerators based on residue number systems~\cite{demirkiran2024blueprint,demirkiran2024mirage}, where implementing many parallel modular channels is infeasible due to optical hardware overhead and noise sensitivity. In all these cases, closed-form optimal constructions for small $L$ are highly desirable.

To address this need, we derive closed-form optimal solutions for \(L = 2, 3,\) and \(4\), which maximize the scaling factor \(m\) under the design constraints \eqref{eq:pairwise_coprime}–\eqref{eq:range_constraint}. These constructions rely on the following elementary lemma on the pairwise coprimality of integers.

\begin{lemma}[Basic Coprimality Properties]
\label{lem:coprime_properties}
Let $\alpha \in \mathbb{N}$. Then the following hold:
\begin{enumerate}
\item [\textit{(i)}] Two consecutive integers $\alpha$ and $\alpha + 1$ are coprime, i.e., $\gcd(\alpha, \alpha+1) = 1$.
\item[(ii)] Two consecutive odd integers $2\alpha - 1$ and $2\alpha + 1$ are coprime.
\item[(iii)] Any three consecutive odd integers $\{2\alpha - 1,\ 2\alpha + 1,\ 2\alpha + 3\}$ are pairwise coprime.
\item[(iv)] Any three consecutive integers $\{\alpha,\ \alpha+1,\ \alpha+2\}$ with $\alpha$ odd are pairwise coprime.
\end{enumerate}
\end{lemma}

\begin{IEEEproof}
The proof is based on the elementary fact that for any two positive integers $a < b$ and $ 1 \le k \le \left\lfloor b/a \right\rfloor$, 
\begin{equation}\label{eq:baic_coprime}
\gcd(b,a)=\gcd(b,b-ka)=\gcd(a,b-ka).   
\end{equation}
In the special case, when $k=1$, this implies that their greatest common divisor $\gcd(a,b)$ must divide the difference $b-a$. 

\emph{\textit{(i)}} As $\gcd(\alpha, \alpha+1)$ divides $1$, we have $\gcd(\alpha, \alpha+1) = 1$.

\emph{(ii)} Let $a = 2\alpha - 1$ and $b = 2\alpha + 1$. Then $\gcd(a,b) = \gcd(a, b - a) = \gcd(a, 2)$.
Since $a$ is odd, $\gcd(a,2) = 1$ and hence $\gcd(a,b) = 1$.

\emph{(iii)} Consider three consecutive odd integers $a = 2\alpha - 1,\ b = 2\alpha + 1,\ c = 2\alpha + 3$.
We have already shown that $\gcd(a,b) = \gcd(b,c) = 1$ in \textit{(ii)}.  
It remains to show $\gcd(a,c) = \gcd(2\alpha - 1, 2\alpha + 3) = \gcd(2\alpha - 1, 4)$.  
As $2\alpha - 1$ is odd, it cannot be divisible by 2 or 4, so the gcd is 1.

\emph{(iv)} Let $\alpha$ be odd. Then $\alpha$ is coprime to both $\alpha+1$ and $\alpha+2$ by part (\emph{i}).
Also, $\gcd(\alpha+1, \alpha+2) = 1$, so all three are pairwise coprime.
\end{IEEEproof}

Based on Lemma~\ref{lem:coprime_properties}, we now present optimal constructions for \( L = 2, 3 \), and \( 4 \), maximizing the feasible value of \( m = m_{\max} / \Gamma_L \).

\begin{theorem}[Optimal Moduli Selection for $L=2,3,4$]
\label{thm:L2L3Optimal}
Let $\rho>1$ be given by~\eqref{eq:rho_def}. Under the design constraints~\eqref{eq:pairwise_coprime}–\eqref{eq:range_constraint}, the following constructions maximize 
$m = m_{\max}/\Gamma_L$.

\begin{itemize}
    \item[$\bullet$] \textbf{Case $L=2$}:  
    The optimal coprime pair is
    \begin{equation}\label{eq:optl2}
        \{ \Gamma_1, \Gamma_2 \} = \{ \lceil \rho \rceil, \ \lceil \rho \rceil + 1 \}.
    \end{equation}

    \item[$\bullet$] \textbf{Case $L=3$}:  
    When  $1<\rho\le 6$, the unique optimal triple is
\[
(\Gamma_1,\Gamma_2,\Gamma_3)=(2,3,5) .
\]   
When $\rho>6$, define
    \begin{equation}\label{eq:defb}
        b = \min\{n\in \mathbb{Z}, \, n \geq 3: n(n-1) \geq \rho \,\} 
    \end{equation}
    Then the optimal triple is chosen as follows:
    \begin{align}
        &\{ \Gamma_1, \Gamma_2, \Gamma_3 \} = \{ b-1,\, b,\, b+1 \}, && \text{if $b$ is even}, \\
        &\{ \Gamma_1, \Gamma_2, \Gamma_3 \} = \{ b-2,\, b,\, b+1 \}, && \text{if $b$ is odd, $b \geq 5$, $(b-2)b \geq \rho$, and $(b+1)\bmod 3 \neq 0$}, \\
        &\{ \Gamma_1, \Gamma_2, \Gamma_3 \} = \{ b,\, b+1,\, b+2 \}, && \text{otherwise}.
    \end{align}

    \item[$\bullet$] \textbf{Case $L=4$:}
    When $1<\rho\le 30$, the unique optimal (sorted, pairwise–coprime) quadruple is
\[
(\Gamma_1,\Gamma_2,\Gamma_3,\Gamma_4)=(2,3,5,7).
\]   
    When $\rho>30$, define
    \begin{equation}\label{eq:defb_L4}
        b = \min\{n\in \mathbb{Z}, \, n \geq 5: n(n-1)(n-2) \geq \rho \,\}.
    \end{equation}
    The corresponding optimal quartets $\{\Gamma_i\}$ are enumerated in Table~\ref{tab:opt4}.
\end{itemize}
\end{theorem}

The proof will be provided in Appendix~A. The constructions in Theorem~\ref{thm:L2L3Optimal} are explicit and involve only a few arithmetic operations and comparisons. Hence, for \(L = 2,3\) and $4$, the optimal moduli can be computed easily, making them suitable for hardware front-ends.

\begin{table*}[t]
\caption{Optimal quartet $\{\Gamma_i\}_{i=1}^{4}$ for $\rho>6$ with $b$ defined in~\eqref{eq:defb_L4}.  
The symbols “$\land$’’ and “$\lor$’’ denote logical “and’’ / “or’’.}
\label{tab:opt4}
\centering
\small
\begin{tabular}{@{}ll@{}}
\toprule
\textbf{Case label \& product gap} & \textbf{Optimal quartet $\{\Gamma_1,\Gamma_2,\Gamma_3,\Gamma_4\}$} \\
\midrule
\multicolumn{2}{@{}l}{\textbf{A – Odd $b$}} \\
A1 : $b \bmod 3 \ne 1$ & $\{b - 2,\ b - 1,\ b,\ b + 2\}$ \\
A2 : $b \bmod 3 = 1$   & $\{b - 2,\ b,\ b + 1,\ b + 2\}$ \\
\midrule
\multicolumn{2}{@{}l}{\textbf{B – Even $b$ with $(b - 3)(b - 1)b \ge \rho$}} \\
B1 : $b \bmod 6 \in \{2, 4\}$ & $\{b - 3,\ b - 1,\ b,\ b + 1\}$ \\
B2 : $b \bmod 6 = 0 \land b \bmod 5 \ne 3$ & $\{b - 3,\ b - 1,\ b + 1,\ b + 2\}$ \\
B3 : $b \bmod 6 = 0 \land b \bmod 5 = 3$ & \\
\quad B3.1: if $(b - 5)(b - 1)(b + 1) \ge \rho$ $\land$ $b \bmod 7 \ne 5$ & $\{b - 5,\ b - 1,\ b + 1,\ b + 2\}$ \\
\quad B3.2: otherwise & $\{b - 1,\ b + 1,\ b + 2,\ b + 3\}$ \\
\midrule
\multicolumn{2}{@{}l}{\textbf{C – Even $b$ with $(b - 3)(b - 1)b < \rho \le (b - 3)(b - 1)(b + 1)$}} \\
C1 : $b \bmod 3 \ne 1 \land b \bmod 5 \ne 3$ & $\{b - 3,\ b - 1,\ b + 1,\ b + 2\}$ \\
C2 : $b \bmod 3 = 1 \land b \bmod 5 \ne 3$   & $\{b - 1,\ b,\ b + 1,\ b + 3\}$ \\
C3 : $b \bmod 5 = 3$                         &  \\
\quad C3.1: $b \bmod 3 \neq 1$ & $\{b - 1,\ b + 1,\ b + 2,\ b + 3\}$ \\
\quad C3.2: $b \bmod 3 = 1$ & $\{b - 1,\ b,\ b + 1,\ b + 3\}$ \\
\midrule
\multicolumn{2}{@{}l}{\textbf{D – Even $b$ with $(b - 3)(b - 1)(b + 1) < \rho$}} \\
D1 : $b \bmod 3 \ne 0 \land b \bmod 5 \ne 3$ & $\{b - 1,\ b,\ b + 1,\ b + 3\}$ \\
D2 : $b \bmod 3 = 0 \lor b \bmod 5 = 3$      & $\{b - 1,\ b + 1,\ b + 2,\ b + 3\}$ \\
\bottomrule
\end{tabular}
\end{table*}

\begin{example} Suppose that \( m_{\max} = 100 \) and \( N_{\text{th}} = 10^6 \), yielding \( \rho =N_{\text{th}}/m_{max}= 10^4 \).

For \(L = 3\), we compute \( b = 101 \) from~\eqref{eq:defb}. As \( b \) is odd and the fallback condition fails because \( (b{-}2)\cdot b < \rho \) and \( (b{+}1) \bmod 3 = 0 \). We therefore adopt the configuration \(\{101, 102, 103\}\), which is pairwise coprime and yields a common factor \( m = 100 / 103 \approx 0.97 \).

For \(L = 4\), we compute \( b = 22 \). The selected set \(\{21, 22, 23, 25\}\) corresponds to \textbf{Case C2} in Table~\ref{tab:opt4}, where \( b \bmod 3 = 1 \) and \( b \bmod 5 \neq 3 \). This yields \( \Gamma_4 = 25 \), giving a common factor \( m = 100 / 25 = 4 \).
\end{example}

\begin{example}
Consider a 3-moduli setting (\( L = 3 \)) with dynamic range threshold \( N_{\text{th}} = 2 \times 10^4 \) and modulus bound \( m_{\max} = 55 \), yielding \( \lceil \rho \rceil = 364 \). We compare the proposed modulus selection strategy against two common alternatives:  
(\emph{i}) restricting all \( \Gamma_i \) to prime numbers~\cite{xiao_robustness_2018}, and  
(\emph{ii}) using the hardware-friendly form \( \{2^n - 1, 2^n, 2^n + 1\} \), commonly used in residue-based data storage~\cite{abdallah2005multimoduli}. Using Theorem~\ref{thm:L2L3Optimal}, our method yields \( (\Gamma_1, \Gamma_2, \Gamma_3) = (19, 20, 21) \) with scaling factor \( m = 2.6190 \). Among prime-only configurations, the best is \( (17, 23, 29) \) with \( m = 1.8966 \), while the best structured triple is \( (31, 32, 33) \), yielding \( m = 1.6667 \). This implies our design has a \( 38\% \) improvement over prime-only methods and a \( 57\% \) improvement over structured digital alternatives.
\end{example}

\noindent\textbf{Hardware-friendly design.} A key advantage of the proposed closed-form construction is its compatibility with finite-precision hardware implementations. In practical systems, the moduli \(m_i\) are typically realized through analog circuits or digitally controlled elements with limited resolution. Our design permits simple precision management: the scalar \(m\) can be truncated (denoted \(\hat{m}\)) to match hardware precision (e.g., \(\ell\) decimal places), while keeping the \(\Gamma_i\) unchanged. This allows the RCRT formula to remain valid, with only slight degradation in dynamic range due to scaling by \(\hat{m}/m\).

Note that some existing works~\cite{akhlaq2016selecting} select rational moduli that are not pairwise coprime, leading to modulus values like $\frac{101039}{66}$, $\frac{1076285}{682}$, etc. (see Example D in~\cite{akhlaq2016selecting}). These require precise handling of both numerators and denominators, potentially necessitating more complex re-optimization of moduli to meet hardware precision limits.

\begin{Rem} For $L = 4$ and even $b$, the expressions for the optimal solution $\{\Gamma_1, \Gamma_2, \Gamma_3, \Gamma_4\}$ in Table~\ref{tab:opt4} can become lengthy due to the need to account for different ranges of $\rho$. To simplify the design while maintaining near-optimality, one may instead consider $b' = b + 1$ and select the candidate set $\{b'-2,\ b'-1,\ b',\ b'+2\}$ when $b' \bmod 3 \ne 1$, or $\{b'-2,\ b'-1,\ b',\ b'+3\}$ when $b' \bmod 3 = 1$ as in \textbf{Case A} of Table~\ref{tab:opt4}. This heuristic yields simplified selection with performance close to optimal.

For systems that can tolerate one or two additional channels (\(L = 5\) or \(6\)), an exact optimization quickly becomes prohibitive, yet a simple heuristic suffices in practice. Starting from the closed-form quartet in Table~\ref{tab:opt4}, one can append each extra modulus by (\emph{i}) estimating the size that would keep the overall product on the order of \(\rho^{1/(L-1)}\), (\emph{ii}) scanning outward for the nearest prime moduli, and (\emph{iii}) updating the cumulative product to calculate the remaining moduli. 
\end{Rem}

\section{Fibonacci-Inspired Two-Moduli Design for Layered RCRT Reconstruction} 
\label{sec:Multi-level}
\subsection{Basics of Fibonacci-like Sequences}\label{sec:basic-fblike}

As a step beyond the full CRT layer moduli selection in Sec.~\ref{sec:opSe}, we now consider the design of two-moduli RCRT systems that support \emph{layered reconstruction}. 
Note that the sequence $\sigma_j$ in~\eqref{eq:sigmaexp_xiao} corresponds to the remainder chain of the Euclidean algorithm computing $\gcd(\Gamma_1,\Gamma_2)$. As $\Gamma_{1}$ and $\Gamma_{2}$ are coprime, there exists a finite index $K$ such that $\sigma_{K+1}=1$. Notably, consecutive Fibonacci numbers $(F_{K+3},F_{K+4})$ represent the smallest inputs requiring exactly $K+1$ Euclidean steps. Leveraging this connection, we introduce Fibonacci-like integer sequences whose algebraic structure guarantees the desired remainder chain.
\begin{deft}[Fibonacci-like sequence \( \{F_{d,k}\} \)]\label{deft:Fib_Seq}
Let \( d \ge 0 \) be an integer. The Fibonacci-like sequence \( \{F_{d,k}\} \) is defined recursively by
\begin{equation}\label{eq:def_fib}
F_{d,0} = d,\:  F_{d,1} = 1, \:F_{d,k}=F_{d,k-1}+F_{d,k-2}\quad (k\ge 2)
 \end{equation}
This extends the classical Fibonacci sequence by initializing with seed \( d \) at $k=0$.
\end{deft}

\begin{Rem}
When \( d = 0 \), the sequence \( \{F_{d,k}\} \) reduces to the standard Fibonacci sequence \( (0, 1, 1, 2, 3, \cdots) \), i.e., \( F_{0,k} = F_k \).  
For \( d = 1 \), it becomes a one-step right shift of the standard sequence: \( F_{1,k} = F_{0,k+1} \).  
When \( d = 2 \), the sequence coincides with the Lucas numbers~\cite{vajda2008fibonacci}. The generalized sequence \( \{F_{d,k}\} \) preserves key number-theoretic properties of the Fibonacci sequence, such as the coprimality of consecutive terms: \( \gcd(F_{d,k}, F_{d,k+1}) = 1 \).  
Since the recurrence relation remains unchanged, i.e., \( F_{d,k} = F_{d,k-1} + F_{d,k-2} \), its asymptotic growth is exponential and governed by the golden ratio \( \varphi \):
\begin{equation}\label{eq:growth}
  \lim_{k \to \infty} \frac{F_{d,k+l}}{F_{d,k}} = \varphi^{l}.  
\end{equation}
\end{Rem}

We next prove two key algebraic properties of \( \{F_{d,k}\} \) instrumental for our recursive construction.

\begin{lemma}[Properties of the Fibonacci-like Sequence \( \{F_{d,k}\} \)]\label{lemma:FibLikeProperties}
Let \( d \in \mathbb{Z}_{\geq0} \), and let \( \{F_{d,k}\} \) denote the Fibonacci-like sequence in Definition~\ref{deft:Fib_Seq}. Then the following hold:
\begin{enumerate}[label=\roman*.]
    \item[\textit{(i)}] \textbf{Linear in seed property}: For all \( k \ge 2 \),
    \begin{equation}
    \label{eq:Fibod}
        F_{d,k} = F_{1,k-1} + d\,F_{1,k-2}.
    \end{equation}
    \item[\textit{(ii)}] \textbf{Mixed d\textquotesingle Ocagne identity}: For \( s,t \in \mathbb{Z} \) and \(s\ge t\ge1 \),
    \begin{equation}
    \label{eq:cassiForFibo}
     F_{d,s} F_{1,t} - F_{d,s+1} F_{1,t-1} = (-1)^t F_{d,s-t}.
    \end{equation}
\end{enumerate}
\end{lemma}

\begin{IEEEproof}
\textit{(i)} We prove by induction on \(k \ge 2\).  
Base case \(k = 2\):  
\(
F_{d,2} = F_{d,1} + F_{d,0} = 1 + d = F_{1,1} + d F_{1,0}.
\)  

Assume the identity holds for \(k\) and \(k-1\). Then:
\begin{align*}
    F_{d,k+1} & = F_{d,k} + F_{d,k-1} \\
    &= \big(F_{1,k-1} + d\,F_{1,k-2}\big) + \big(F_{1,k-2} + d\,F_{1,k-3}\big) \\
    &= F_{1,k-1} + F_{1,k-2} + d\,\big(F_{1,k-2} + F_{1,k-3}\big)\\
    &=F_{1,k}+dF_{1,k-1}.
\end{align*}
Hence, by the recurrence of the Fibonacci sequence, the identity holds for all \(k \ge 2\).

\textit{(ii)}  By the classical d\textquotesingle Ocagne's identity~\cite{Koshy2018,Keskin2014ThreeIdentities}, for $a,b\in \mathbb{Z}_{>0}$ and ($a\ge b$), the Fibonacci sequence $F_{0,k}$ satisfies 
\begin{equation}\label{eq:fibd0}
F_{0,a+1} F_{0,b+1} - F_{0,a+2} F_{0,b} = (-1)^b F_{0,a-b+1},
\end{equation}
Since $F_{1,k}=F_{0,k+1}$, this implies 
\begin{equation}\label{eq:fibd1}
F_{1,a} F_{1,b} - F_{1,a+1} F_{1,b-1} = (-1)^b F_{1,a-b}.
\end{equation}
That is,~\eqref{eq:cassiForFibo} holds for $d=1$. When $d>1$, applying~\eqref{eq:Fibod} for $F_{d,s}$ and $F_{d,s+1}$, the left-hand side (LHS) of~\eqref{eq:cassiForFibo} can be written as 
\begin{align*}
& F_{d,s} F_{1,t} - F_{d,s+1} F_{1,t-1} \\
= &(F_{1,s-1} + d\,F_{1,s-2}) F_{1,t} - (F_{1,s} + d\,F_{1,s-1}) F_{1,t-1} \\
= &[F_{1,s-1} F_{1,t} - F_{1,s} F_{1,t-1}] 
+ d [F_{1,s-2} F_{1,t} - F_{1,s-1} F_{1,t-1}].
\end{align*}
Using~\eqref{eq:fibd1}, the above can be simplified to 
\[
(-1)^t F_{1,s-t-1} + d(-1)^t F_{1,s-t-2} = (-1)^t F_{d,s-t},
\]
which completes the proof.
\end{IEEEproof}

\subsection{Construction of Moduli for Layered-Design}
We now formalize the two–moduli design for \emph{layered RCRT}. Select two moduli
\[
\{m_1,m_2\}=\{m\,\Gamma_1,\; m\,\Gamma_2\},\qquad \Gamma_1<\Gamma_2,
\]
to support one full CRT layer and $K$ additional robust layers while satisfying
Eqs.~\eqref{eq:pairwise_coprime}–\eqref{eq:range_constraint}. Our goal is to
\emph{maximize} the guaranteed full CRT layer error tolerance, which is achieved by maximizing $m$. Since the modulus bound implies
$m=m_{\max}/\Gamma_2$, this is equivalent to minimizing $\Gamma_2$.
Accordingly, the design problem is
\begin{equation}
\label{eq:layered_obj}
\min_{\Gamma_1,\Gamma_2\in\mathbb{Z}}\ \Gamma_2
\end{equation}
subject to
\begin{align}
\Gamma_2\ >\ \Gamma_1\ \ge\ \rho\ >\ 1,\quad & \Gamma_\ell\in\mathbb{Z}\ (\ell=1,2),
\label{eq:range_constraint_layered}\\
\gcd(\Gamma_1,\Gamma_2)&=1, \label{eq:coprime_layered}\\
\sigma_{K+1}&=1, \label{eq:depth_constraint}
\end{align}
where $\rho$ and the remainder chain $\{\sigma_k\}$ are defined in
\eqref{eq:rho_def} and \eqref{eq:sigmaexp_xiao}, respectively. Minimizing
$\Gamma_2$ under \eqref{eq:range_constraint_layered}–\eqref{eq:depth_constraint}
therefore maximizes $m$ and hence the decoder’s full CRT layer error tolerance, while
ensuring exactly $K$ robust layers.

A key fact is that the consecutive Fibonacci pair \((F_{1,K+2}, F_{1,K+3})\) is the smallest positive integer pair whose Euclidean algorithm takes exactly \(K{+}1\) divisions. Hence, among all pairs that realize the required depth \(\sigma_{K+1}=1\), a consecutive Fibonacci choice minimizes \(\Gamma_2\), provided that the range constraint \(\Gamma_1 \ge \rho\) is already met. This leads to an immediate solution when \(\rho \le F_{1,K+2}\), as stated in the following proposition.

\begin{proposition}\label{prop:fibolargeK}
Fix \(K\ge 1\). If \(\rho \le F_{1,K+2}\), then an optimal solution of
\eqref{eq:layered_obj} subject to
\eqref{eq:range_constraint_layered}--\eqref{eq:depth_constraint} is
\[
\Gamma_1 = F_{1,K+2},\qquad \Gamma_2 = F_{1,K+3}.
\]
A convenient estimate for the smallest layer index \(K^\ast\) with
\(\rho \le F_{1,K^\ast+2}\) is
\begin{equation}
\label{eq:Kstar}
K^\ast \;=\; \max\!\left\{\,1,\;\left\lceil \frac{\ln(\rho\sqrt{5})}{\ln\varphi} - 3 \right\rceil \right\},
\qquad \varphi=\tfrac{1+\sqrt{5}}{2}.
\end{equation}
\end{proposition}

\begin{IEEEproof}
A classical fact on Euclidean algorithm (see, e.g., \cite{shallit1994origins})
is that the smallest positive integers yielding exactly \(K{+}1\) divisions and
terminating at \(1\) are consecutive Fibonacci numbers \((F_{0,K+3},F_{0,K+4})\).
Using \(F_{1,k}=F_{0,k+1}\) gives \((\Gamma_1,\Gamma_2)=(F_{1,K+2},F_{1,K+3})\),
which minimizes \(\Gamma_2\) among all pairs meeting the depth constraint
\(\sigma_{K+1}=1\) and \(\Gamma_1\ge\rho\). For \eqref{eq:Kstar}, apply Binet’s
approximation \(F_{1,k}\approx \varphi^{k+1}/\sqrt{5}\) and solve
\(F_{1,K^\ast+2}\ge \rho\) for \(K^\ast\).
\end{IEEEproof}


Theorem~\ref{thm2:miniGamma} provides a constructive
solution when $\rho>F_{1,K+2}$.

\begin{theorem}[Fibonacci-Inspired Moduli Construction]
\label{thm2:miniGamma}
Let $\{F_{d,k}\}$ be the $d$-seed Fibonacci sequence given in Definition~\ref{deft:Fib_Seq}. 
Fix $K\!\ge\!1$ and let $\rho=N_{\text{th}}/m_{\max}$ be given with $\rho>F_{1,K+2}$.
For each integer
seed $d\ge 1$, set
\begin{gather}
\zeta_d = \lceil(\rho - F_{d,K})/F_{d,K+1}\rceil, \label{eq:zetad_def} \\
\Gamma_1(d) = \zeta_d F_{d,K+1} + F_{d,K}, \ \Gamma_2(d) = \Gamma_1(d) + F_{d,K+1}. \label{eq:gamma12-d}
\end{gather}
\noindent\textbf{\textit{(i)} Remainder Chain and Dynamic Range:}
The pair $(\Gamma_1(d),\Gamma_2(d))$ is coprime with $\Gamma_1(d) \geq \rho$, having Euclidean remainders:
\begin{equation}\label{eq:defsigmajthm}
\sigma_j = F_{d,K+2-j}, \quad 1 \leq j \leq K+1.
\end{equation}
The dynamic-range breakpoints $P_j$ ($1 \leq j \leq K$) are:
\begin{equation}\label{eq:Pj_compact}
P_j = \begin{cases}
m\Gamma_1(d) F_{\zeta_d,2j_0}, & j=2j_0-1, \\[3pt]
m\Gamma_2(d) F_{\zeta_d-1,2j_0+1}, & j=2j_0,
\end{cases}
\end{equation}
with $P_{K+1} = \Gamma_1(d)\Gamma_2(d)m$. Robust recovery holds for $x \in [0,P_j)$ when $| e_1 - e_2| < m\sigma_j/2$.

\noindent\textbf{(ii) Optimal Seed Selection:}
For $\Gamma_1(d)$ and $\Gamma_1(d)$ defined in~\eqref{eq:gamma12-d},  the minimal $\Gamma_2$ is obtained via:
\[
(\Gamma_1^\star, \Gamma_2^\star) = \arg\min_{1 \leq d \leq 3} \Gamma_2(d),
\]
where testing $d=1,2,3$ suffices, i.e., no larger seed yields smaller $\Gamma_2$ in \eqref{eq:gamma12-d}, thus maximizing $m = m_{\max}/\Gamma_2^\star$.
\end{theorem}

The proof is in Appendix~B. 
\begin{Rem}
  When $\rho \leq F_{1,K+2}$, $(F_{1,K+2}, F_{1,K+3})$ is optimal (no search is needed). For $\rho > F_{1,K+2}$, each seed $d$ defines $\zeta_d$ via~\eqref{eq:zetad_def} to ensure $\Gamma_1(d) \geq \rho$, yielding candidate pairs $(\Gamma_1(d),\Gamma_2(d))$. Theorem~\ref{thm2:miniGamma} guarantees that the family-optimal pair is found by evaluating only $d = 1, 2, 3$ via~\eqref{eq:zetad_def} and~\eqref{eq:gamma12-d}, requiring just three comparisons. 
\end{Rem}

Corollary~\ref{cor:explicit_opt_pairs} next shows that when $K \leq 2$, this construction actually yields \emph{globally optimal} solutions with deterministic seed selection of $d$.

\begin{corollary}[Closed-Form Global Optima for $K \leq 2$]
\label{cor:explicit_opt_pairs}
For $\rho>F_{1,K+2}$ and $K \in \{1,2\}$, Theorem~\ref{thm2:miniGamma}'s construction achieves global optimality with:
\begin{enumerate}
\item[\textit{(i)}] \textbf{$K=1$}: Always use $d=1$ with
\[
\zeta_1 = \lceil (\rho - 1)/2 \rceil,\quad 
\Gamma_1 = 2\zeta_1 + 1,\quad 
\Gamma_2 = 2\zeta_1 + 3
\]

\item[(ii)] \textbf{$K=2$}: For $\lceil \rho\rceil  = 12\ell + 3$ ($\ell\ge 1$), take $d=2$:
\[
\zeta_2 = 3\ell,\quad 
\Gamma_1 = 4\zeta_2 + 3,\quad 
\Gamma_2 = 4\zeta_2 + 7
\]
Otherwise, use $d=1$:
\[
\zeta_1 = \lceil (\rho - 2)/3 \rceil,\quad 
\Gamma_1 = 3\zeta_1 + 2,\quad 
\Gamma_2 = 3\zeta_1 + 5
\]
\end{enumerate}
In all cases, $\Gamma_2$ is globally minimized (maximizing $m$).
\end{corollary}
The proof is given in Appendix~C. 

\begin{example}
Table~\ref{Table:Comparison1} illustrates the construction for various values of \(\rho\) and \(K\). Although Theorem~\ref{thm2:miniGamma} requires checking \(d = 1, 2, 3\) for optimality, the case \(d = 1\) alone yields near-optimal results across all tested scenarios, achieving an excellent trade-off between performance and simplicity.
\end{example}

\begin{table}[t]
\centering
\caption{Examples of ($\Gamma_1, \Gamma_2$) under different $\rho$ and $K$}
\renewcommand{\arraystretch}{1.5}
\begin{tabular}{ccccccc}
\toprule
\toprule
$\rho$ & $K$ & $d$ & $\Gamma_1$ & $\Gamma_2$ & $\sigma_j$: $\sigma_1,\,\sigma_2,\,\dots,\,\sigma_{K+1}$ & $\zeta_d$\\
\midrule
\multirow{5}{*}{150}   & 1 & 1 & 151 & 153 & $(2,\,1)$ & 75
\\
\cline{2-7}
& \multirow{1}{*}{2} & 1 & 152 & 155 & $(3,\,2,\,1)$ & 50\\
\cline{2-7}
& \multirow{1}{*}{3} & 1 & 153 & 158 & $(5,\,3,\,2,\,1)$ & 30 \\
\cline{2-7}
& \multirow{2}{*}{4} & 1 & 157 & 165 & $(8,\,5,\,3,\,2,\,1)$ & 19 \\
\cline{3-7}
& & 3 & 150 & 161 & $(11,\,7,\,4,\,3,\,1)$ & 13 \\
\midrule

\multirow{2}{*}{250} & \multirow{2}{*}{6}  & 1 & 265 & 286 & $(21,\,13,\,8,\,5,\,3,\,2,\,1)$ & 12 \\
\cline{3-7}

& & 2 & 250 & 279 & $(29,\,18,\,11,\, 7, \, 4,\,3,\,1)$ & 8 \\
\bottomrule
\bottomrule
\end{tabular}\label{Table:Comparison1}
\end{table}

\begin{Rem}
Corollary~\ref{cor:explicit_opt_pairs} guarantees global optimality for \(K \leq 2\), but \emph{not} for \(K \geq 3\). For instance, when \(K = 3\) and \(\rho = 19\), Theorem~\ref{thm2:miniGamma} with \(d = 1\) yields the near-optimal pair \((23, 28)\), while \((19, 26)\) has a smaller \(\Gamma_2\) with the same \(K+1 = 4\) steps. Remarkably, even in this case, the \(d = 1\) construction stays within \(8\%\) of the optimal \(\Gamma_2\). Although global optimality for \(K \geq 3\) remains open, the method's efficiency and consistently near-optimal performance make it practically attractive.
\end{Rem}

\section{Layer-Wise Performance Analysis: Scaling Laws and Statistical Behavior}\label{sec:Layered}
\subsection{Layer-Wise Scaling of Dynamic Range and Error Bounds}
To understand how the number of robust layers \(K\) influences the system's performance, we theoretically analyze the dynamic range and error tolerance provided by our layered construction. This cross-layer analysis characterizes how both dynamic range and error tolerance scale with depth, covering the full CRT layer as well as intermediate robust layers, and reveals the trade-offs inherent in choosing $K$. 

\textbf{Full CRT Layer:} From Sec.~\ref{sec:opSe}, we recall the moduli definitions:
$
m_1 = m\,\Gamma_1,\quad m_2 = m\,\Gamma_2 = m_{\max},
$ where \(m_{\max}\) is the hardware-constrained largest modulus. The scaling factor is therefore \(m = m_{\max} / \Gamma_2\), and for the final reconstruction layer (\(j = K{+}1\)), we have:
\begin{equation}
\tau_{K+1} = \frac{m}{4} = \frac{m_{\max}}{4\,\Gamma_2},
\quad
P_{K+1} = m\,\Gamma_1\Gamma_2 = \Gamma_1\,m_{\max}.
\label{eq:PK+1_delta}
\end{equation}

\paragraph*{1) Exponential Scaling Regime}
 From proposition~\ref{prop:fibolargeK}, we know that when \(K > K^*\), the optimal pair \((\Gamma_1, \Gamma_2) = (F_{1,K+2}, F_{1,K+3})\). By~\eqref{eq:PK+1_delta}, $P_{K+1}\propto \Gamma_1$ and $\tau_{K+1}\propto \Gamma^{-1}_2$, which implies that 
\[
P_{K+1} \propto \varphi^{K},\quad \tau_{K+1} \propto \varphi^{-K},
\]
illustrating a controlled trade-off between dynamic range and robustness depth.

\paragraph*{2) Small-Depth Regime (\(K < 3\))}
When \(K=0\), we recover the classical CRT with adjacent coprime moduli: \(\Gamma_1 = \lceil\rho\rceil\), \(\Gamma_2 = \Gamma_1 + 1\). For \(K=1,2\), Corollary~\ref{cor:explicit_opt_pairs} provides closed-form constructions with \(\Gamma_i\) differing from the \(K=0\) baseline by small offsets, yielding similar dynamic range and noise tolerance at full CRT layer.

\textbf{Intermediate Layers:} Theorem~\ref{thm2:miniGamma} indicates that the error tolerance bound $T(x)$ as a function of signal amplitude \(x\) follows a staircase profile 
\begin{equation} 
T(x) = \tau_j \quad \text{if} \quad P_{j-1} \le x < P_j, \qquad P_0 := 0.
\label{eq:layerErrorDy}
\end{equation}
in which $P_j$ takes the form of ~\eqref{eq:Pj_compact} and the per-residue error bound $\tau_j$ is given by 
\begin{equation}\label{eq:deftauj}
    \tau_j = \frac{m}{4}\sigma_j=\frac{m}{4}F_{d,K+2-j}.
\end{equation}
The next Corollary
describes how $P_j$ and $\tau_j$ ($1\le j \le K$) evolve acroos 
$K$ robustness layers.

\begin{corollary}[Scaling Laws for Intermediate Layers]\label{corl:intermediate}
For the layered construction in Theorem~\ref{thm2:miniGamma} with $K\ge 1$:
\begin{itemize}
\item Error bounds converge as:
     \begin{equation}\label{eq:sigmaj_layer}
        \frac{\tau_j}{\tau_{j+1}} = \frac{F_{d,K+2-j}}{F_{d,K+1-j}} \to \varphi 
        \quad (j\text{ fixed}, K\to\infty)
      \end{equation}

\item Plateaux ratios satisfy:
      \begin{gather}
        P_2/P_1 = \Gamma_2/\Gamma_1,\quad K\ge 2 \label{eq:first2layer} \\
          P_1/P_{K+1} \approx 1/F_{d,K+1},\quad \zeta_d\gg 1 \label{eq:firstlast} \\
        P_K/P_{K+1} < 1/(d+1) \label{eq:robust_bound}\\
              2 < P_{j+2}/P_j < 3\quad (1\le j \le K-2,\:K\ge 3) \label{eq:acrosstwo} \\
        P_{j+2}/P_j \to \varphi^2\approx 2.618 \quad (j\to\infty) \label{eq:acrosstwo_limit} 
      \end{gather}
\end{itemize}
\end{corollary}

\begin{IEEEproof}

(\emph{i}) Eq.~\eqref{eq:sigmaj_layer} is straightforward from ~\eqref{eq:deftauj} and \eqref{eq:growth}.
   
(\emph{ii}) When $K\ge 2$, for the first two robust layers,~\eqref{eq:Pj_compact} implies
\begin{equation}\label{eq:P1P2}
P_{1}=m\Gamma_1F_{\zeta_d-1,3},\quad
P_{2}=m\Gamma_2F_{\zeta_d,2}.
\end{equation}
By definition of Fibonacci-like sequence in~\eqref{eq:def_fib}, one can calculate that $F_{\zeta_d-1,3}=F_{\zeta_d,2}=\zeta_d+1$, indicating~\eqref{eq:first2layer} holds for all $K\ge 2$.   
 
(\emph{iii}) Combining the closed form of \(P_1\) in
\eqref{eq:P1P2} with $F_{\zeta_d-1,3}=\zeta_d+1$ and $P_{K+1}=m\Gamma_1\Gamma_2$ produces
\begin{equation}
    \frac{P_{1}}{P_{K+1}}
       =\frac{\zeta_d+1}
             {(\zeta_d+1)F_{d,K+1}+F_{d,K}}
       \approx \frac{1}{F_{d,K+1}},
\end{equation}
where the approximation $\approx$ assumes $(\zeta_d + 1)\,F_{d,K+1} \gg F_{d,K}$ for typical $\zeta_d\gg 1$ or large $K$.

\textit{(iv)} Substituting $j=K{+}1$ into \eqref{eq:Nj_xiao_alt} gives
\[
N_{K+1}
= N_{K-1} + \Big\lfloor \frac{\sigma_K}{\sigma_{K+1}} \Big\rfloor (N_K - \sigma_K).
\]
Using $\sigma_j=F_{d,K+2-j}$, we have $\sigma_K=F_{d,2}=d{+}1$ and $\sigma_{K+1}=1$, hence
\[
N_{K+1}=N_{K-1} + (d{+}1)(N_K-1).
\]
Since $\{N_k\}$ is monotonically increasing in $k$, we have $N_{K-1}\!\ge\!N_0=\Gamma_2$ for $K\!\ge\!1$.
From \eqref{eq:gamma12-d}, $\Gamma_2>\!F_{d,K+1}\!\ge\!F_{d,2}=d{+}1$, so $N_{K-1}>d{+}1$ and therefore
\[
N_{K+1}
= N_{K-1} + (d{+}1)(N_K-1)
> (d{+}1) + (d{+}1)(N_K-1)
= (d{+}1)N_K.
\]
Recalling $P_{K+1}=mN_{K+1}$ and $P_K=mN_K$, it follows that
\(
P_{K+1}>(d{+}1)P_K,
\)
which proves \eqref{eq:robust_bound}.

\textit{(v)}
From \eqref{eq:Pj_compact} and the recurrence
$F_{d,k+2}=F_{d,k+1}+F_{d,k}=2F_{d,k}+F_{d,k-1}$, we obtain
\[
\frac{P_{j+2}}{P_j} =
\begin{cases}
\dfrac{F_{\zeta_d,\,2j_0+2}}{F_{\zeta_d,\,2j_0}}
= 2 + \dfrac{F_{\zeta_d,\,2j_0-1}}{F_{\zeta_d,\,2j_0}}, & j=2j_0-1, \\[10pt]
\dfrac{F_{\zeta_d-1,\,2j_0+3}}{F_{\zeta_d-1,\,2j_0+1}}
= 2 + \dfrac{F_{\zeta_d-1,\,2j_0}}{F_{\zeta_d-1,\,2j_0+1}}, & j=2j_0.
\end{cases}
\]
For $d\!\ge\!1$ and $k\ge 1$, the Fibonacci-like sequence is strictly increasing in $k$,
i.e, $0<F_{d,k}/F_{d,k+1}<1$. Hence
\(
2 \;<\; P_{j+2}/P_J\;<\; 3,
\)
which proves \eqref{eq:acrosstwo}. Finally, using the above expression for
$P_{j+2}/P_j$ together with the growth law \eqref{eq:growth}
(for which $F_{d,k+2}/F_{d,k}\!\to\!\varphi^2$ as $k\!\to\!\infty$),
we obtain~\eqref{eq:acrosstwo_limit}.
\end{IEEEproof}

\begin{example}
Table~\ref{tab:scaling_laws_verification} evaluates Corollary~\ref{corl:intermediate} for a layered RCRT system with \(\Gamma_1 = 34\), \(\Gamma_2 = 47\) (obtained with $d=1$), and recursion depth \(K = 5\). The ratios \(\sigma_1/\sigma_2 = 1.625\) and \(\sigma_2/\sigma_3 = 1.6\) closely approach \(\varphi\), verifying~\eqref{eq:sigmaj_layer}. The dynamic range ratio \(P_1/P_{K+1} = 0.0638\) agrees with \(1/F_{1,6} = 1/13 \approx 0.0769\) in~\eqref{eq:firstlast}, while \(P_K/P_{K+1} = 0.383 < 1/(d+1) = 0.5\), confirming~\eqref{eq:robust_bound}.
\end{example}

\begin{table}[t]
\centering
\caption{Verification of Scaling Laws in Layered RCRT for  $\Gamma_1=34$ and $\Gamma_2=47.$}
\label{tab:scaling_laws_verification}
\renewcommand{\arraystretch}{1.5}
\begin{tabular}{ccccccc}
\toprule
\hline
$j$ & \textbf{\( \sigma_j \)} & $\sigma_j/\sigma_{j+1}$ & \( P_j \) & \( P_{j+2}/{P_j} \) \\
\hline
1  & \( 13 \)  & $\sigma_1/\sigma_{2}=1.625$ & 102 & \( P_{3}/{P_1} = 2.356\) \\
\hline
2  & \( 8 \)  & $\sigma_2/\sigma_{3}=1.6$ & 141 & \( P_{4}/{P_2} = 2.667\) \\
\hline
3  & \( 5 \)  & $\sigma_3/\sigma_{4}=1.667$ & 238 & \( P_{5}/{P_3} = 2.57\) \\
\hline
4  & \( 3 \)  & $\sigma_4/\sigma_{5}=1.5$ & 376 & - 
\\
\hline
5  & \( 2 \)  & $\sigma_5/\sigma_{6}=2$ &612& - \\
\hline
6  & \( 1 \)  & - &1598 & - \\
\bottomrule
\bottomrule
\end{tabular}
\end{table}

\begin{figure}[!t]
\centering
\subfloat[]{\includegraphics[width=0.7\linewidth]{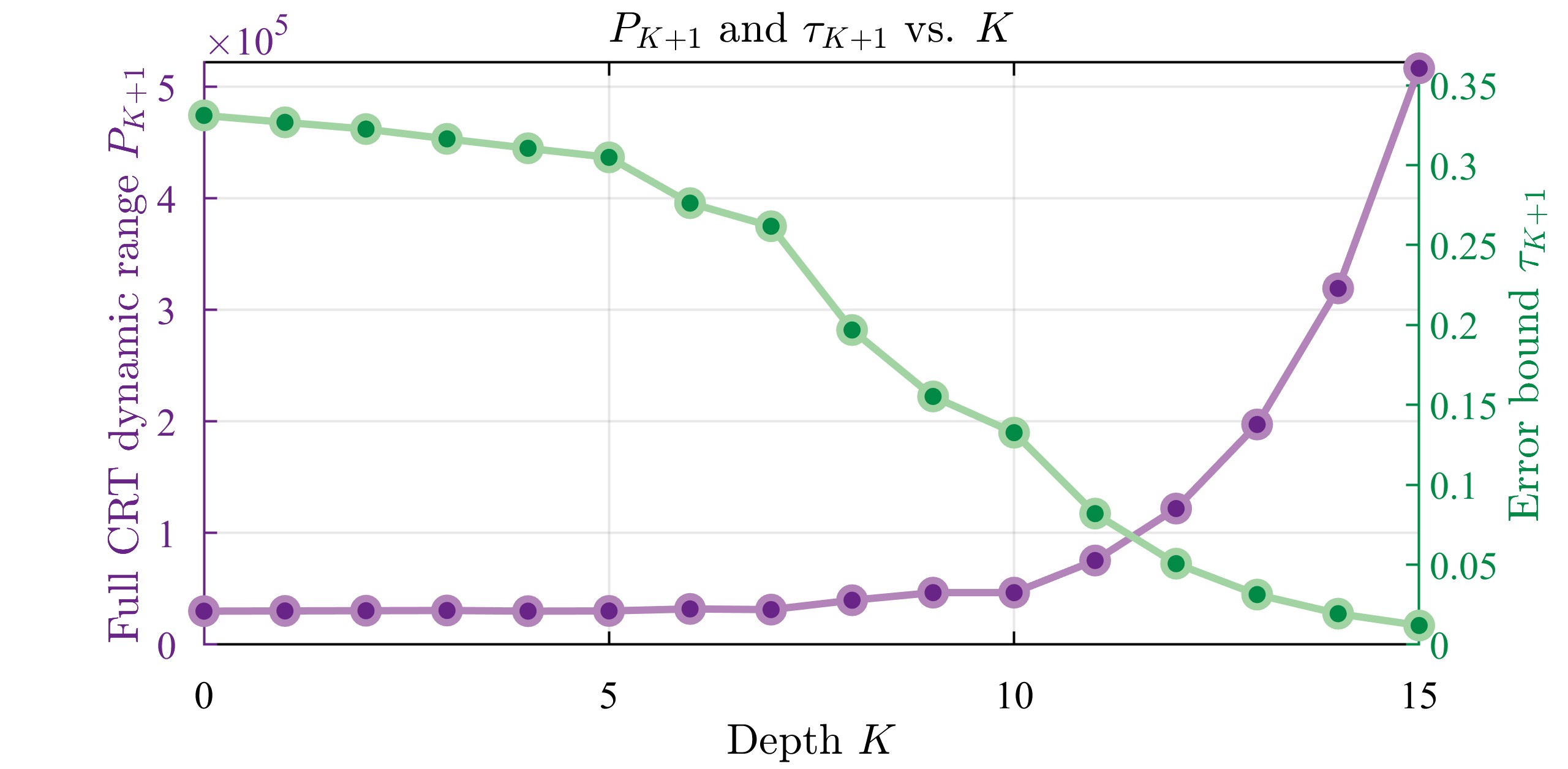}%
\label{fig:LayerWithMP}}\\[1ex]
\subfloat[]{\includegraphics[width=0.7\linewidth]{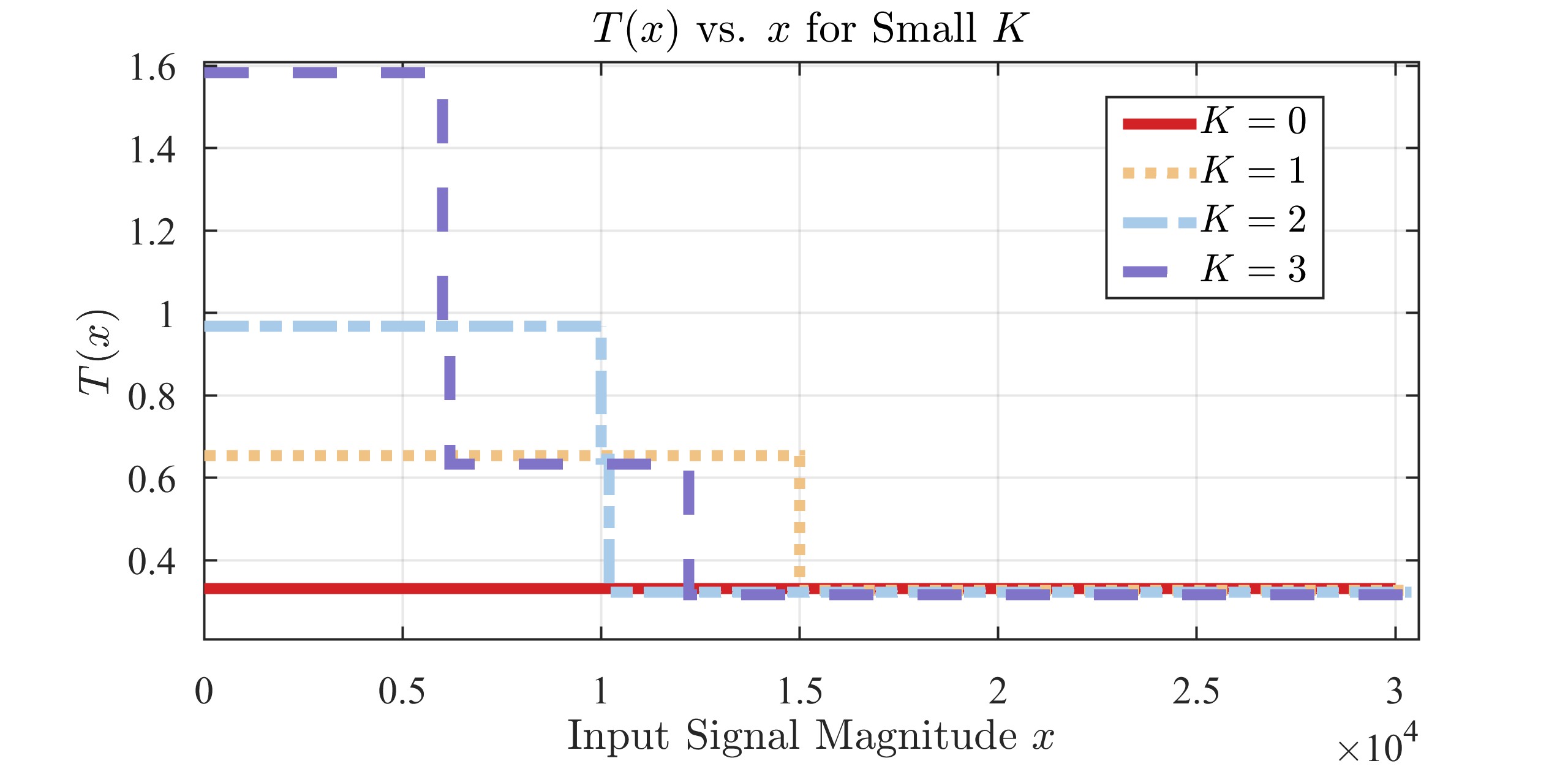}%
\label{fig:KVsm}}\\[1ex]
\subfloat[]{\includegraphics[width=0.7\linewidth]{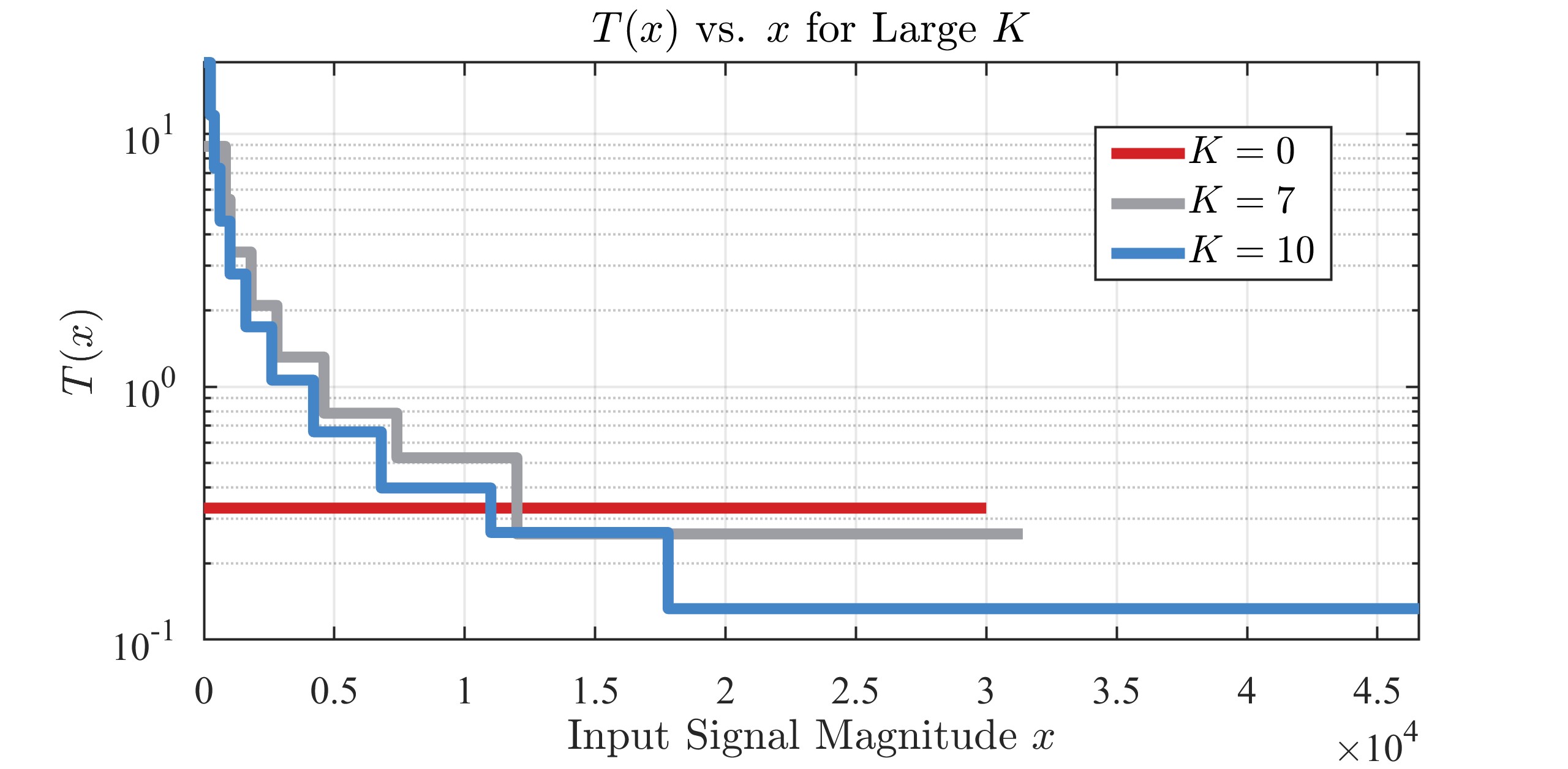}%
\label{fig:KVsP}}
\caption{Error bound vs. dynamic range for the proposed \(K\)-layer robust system. 
\textbf{(a)} Full CRT layer: \(P_{K+1}\) and \(\tau_{K+1}\) vs \(K\). 
\textbf{(b)} $T(x)$ vs. $x$ for \(K = 0, 1, 2, 3\). 
\textbf{(c)} $T(x)$ vs. $x$ for \(K =0, 7, 10\).}
\label{fig:DK}
\vspace{-0.5cm}
\end{figure}

\begin{example}
Given $N_{\text{th}} = 3000$ and $m_{\max} = 200$ (i.e., $\rho = 150$), we use Theorem~\ref{thm2:miniGamma} to construct $(\Gamma_1, \Gamma_2)$ for $0 \le K \le 15$.

Fig.~\ref{fig:LayerWithMP} plots $P_{K+1}$ and $\tau_{K+1}$ vs. $K$.
The critical depth from~\eqref{eq:Kstar} is $K^{\!*}=10$.
For $K\ge K^{\!*}$ the curves follow the predicted
$P_{K+1}\!\propto\!\varphi^{K}$ and
$\tau_{K+1}\!\propto\!\varphi^{-K}$. On the other hand, for $K\le3$ they remain
similar to the $K=0$ baseline, showing that one or two robust
layers incur nearly no full-range loss.

Fig.~\ref{fig:KVsm} ($K\le3$) and Fig.~\ref{fig:KVsP} ($K>3$) depict the
staircase error bound $T(x)$ of~\eqref{eq:layerErrorDy}.
Each step spans $[P_{j-1},P_j)$ and has height
$\tau_j=\frac{m}{4}\sigma_j$. Successive ratios satisfy
$\tau_j/\tau_{j+1}\!\approx\!\varphi$ and
$2<P_{j+2}/P_j<3$.  
As can be observed, small $K$ yields wide plateaux, whereas large $K$ produces a
comb-like staircase whose ever-narrower steps cluster near the origin,
visually confirming the range–robustness trade-off in Corollary~\ref{corl:intermediate}.
\end{example}

\begin{figure}
    \centering
    \includegraphics[width=0.35\linewidth]{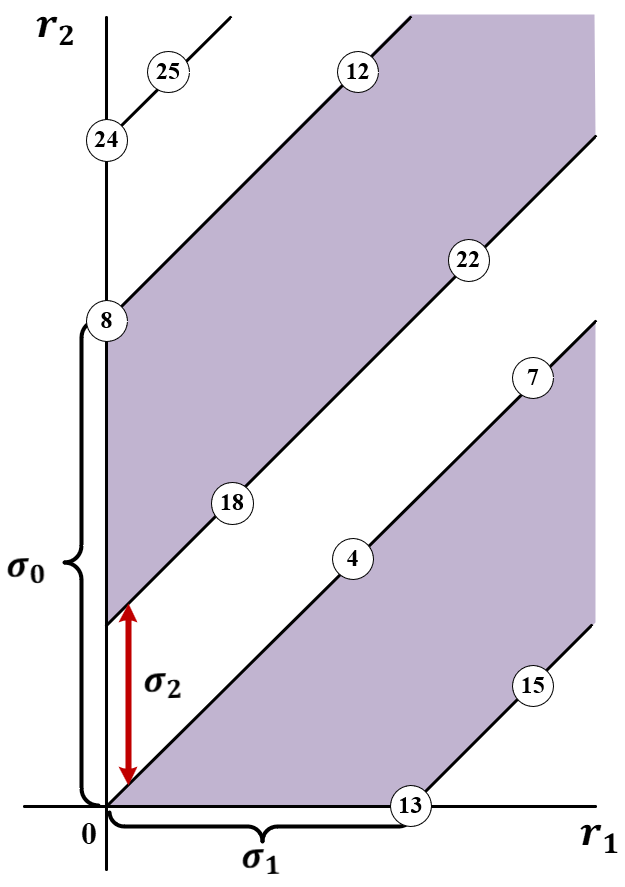}
\caption{Residue space \((r_1, r_2)\) for moduli pair \((m_1, m_2) = (8, 13)\) with \(K = 3\) robust layers. The diagram shows the second layer (\(j = 2\)), where the theoretical minimum line spacing is \(\sigma_2 = 3\), and the dynamic range is \(x\in [0, 26)\). Each slanted line represents a unique folding pair \((n_1, n_2)\), forming a valid decoding region. The shaded area exhibits a gap of \(\sigma_0 - \sigma_2 =\sigma_1=5 > \sigma_2\), revealing non-uniform spacing caused by the residue space geometry.}
    \label{fig:non_uniform}
\end{figure}

\subsection{Statistical Effects on Reconstruction}
In this subsection, we present a statistical analysis of the layered reconstruction performance under noisy residue observations. The goal is to evaluate the probability of successful reconstruction when the input signal $x$ is drawn from a continuous distribution and the observed remainders are corrupted by independent additive noise.
We assume that $x$ is a non-negative real-valued signal with distribution supported on $[0, P_{K+1})$ with very high probability (e.g., probability $\geq 1-10^{-5}$). 

To quantify decoding performance under noise, we follow the layered reconstruction strategy of~\cite{xiao_towards_2017,xiao_robustness_2018}, where the estimate $\tilde{r}_1 - \tilde{r}_2$ is matched to the nearest slanted line associated with valid folding integers. Decoding succeeds in layer $j$ if \textit{(i)} the signal $x$ lies in the dynamic range interval $[P_{j-1}, P_j)$, and \textit{(ii)} the remainder difference satisfies $|e_1 - e_2| < m\sigma_j/2$. This yields a lower bound on the success rate as given below.

\begin{proposition}[Lower bound on success probability (pairwise-difference decoder)]
\label{prop:success_rate}
Consider two-moduli RCRT with $m_\ell=m\,\Gamma_\ell$ ($\ell=1,2$), where $\Gamma_1<\Gamma_2$ are coprime, yielding $K$ robust layers followed by one full CRT layer. Let $x$ be a real-valued random signal drawn from a continuous distribution and assume $x$ is independent of the noise terms $e_1,e_2$, which are independent of each other. Let $\{P_j\}_{j=0}^{K+1}$ be the dynamic-range breakpoints with $P_0=0$ (as in~\eqref{eq:Nj_xiao_alt}), let $\{\sigma_j\}_{j=1}^{K+1}$ be the Euclidean-remainder sequence from~\eqref{eq:sigmaexp_xiao}, and define the layerwise tolerance 
$
\tau_j \;=\; \frac{m\,\sigma_j}{4}\:(1\le j\le K+1).
$
Then the reconstruction success probability satisfies
\begin{equation}
\label{eq:success_prob_general}
\eta_{\mathrm{succ}}
\;\ge\;
\sum_{j=1}^{K+1}
\mathbb{P}\!\left(\,|e_1-e_2| < 2\,\tau_j\,\right)\;
\mathbb{P}\!\left(\,x \in [P_{j-1},P_j)\,\right),
\end{equation}
where $\mathbb{P}(\cdot)$ denotes probability; the first factor is taken with respect to the noise variables $(e_1,e_2)$ and the second with respect to $x$. 
\end{proposition}

\begin{IEEEproof}
On the event $\{x\in [P_{j-1},P_j)\}$, the closed-form pairwise-difference decoder succeeds whenever 
$|e_1-e_2|<\frac{m\sigma_j}{2}=2\tau_j$ (cf. Eq.~\eqref{eq:layered_error_Delta_xiao}). 
Taking the union over disjoint layer intervals and using independence of $x$ and $(e_1,e_2)$ yields
\eqref{eq:success_prob_general} by the law of total probability. 
\end{IEEEproof}

\paragraph*{Typical Signal Priors} 
In many RCRT applications, the input \(x\) follows a known prior distribution, depending on the sensing or communication context. For example, phase retrieval or frequency estimation problems often assume uniform distribution, while communication or radar signals processed by modulo ADCs may follow Rayleigh, exponential, or folded-Gaussian distributions. The interval probability
\(\mathbb{P}(x\!\in\!(P_{j-1},P_j])\) admits a closed-form expression:

\medskip
\begin{itemize}
\item \textbf{Uniform} ($x\!\sim\!\mathcal{U}(0,P_{K+1})$):  
      \(\displaystyle \frac{P_j-P_{j-1}}{P_{K+1}}\).

\item \textbf{Rayleigh} ($x\!\sim\!\mathrm{Rayleigh}(\beta)$, typical for envelope of complex Gaussian signals):  
      \(\displaystyle
        \exp\!\left(-\tfrac{P_{j-1}^2}{2\beta^2}\right) - 
        \exp\!\left(-\tfrac{P_j^2}{2\beta^2}\right)\).

\item \textbf{Exponential} (\(x \sim \text{Exp}(\lambda)\)):  
      \(\displaystyle
        \exp\!\left(-\tfrac{P_{j-1}}{\lambda}\right) -
        \exp\!\left(-\tfrac{P_j}{\lambda}\right)\).

\item \textbf{Folded–Gaussian} (magnitude of real Gaussian signals, $x\sim |\mathcal{N}(0,\vartheta^2)|$):  
      \(\displaystyle
        \operatorname{erf}\!\left(\tfrac{P_j}{\sqrt{2}\vartheta}\right) -
        \operatorname{erf}\!\left(\tfrac{P_{j-1}}{\sqrt{2}\vartheta}\right)\).
\end{itemize}
\medskip

\paragraph*{Typical Noise Models}
Here, we consider two typical noise models.

\textbf{Gaussian Noise:} $e_i \sim \mathcal{N}(0, \varsigma^2)$ independently.
In this case,
\begin{equation} \label{eq:gaussian_prob}
\mathbb{P}(|e_1-e_2| < 2\tau_j) = \text{erf}\left(\frac{\tau_j}{\sqrt{2} \varsigma} \right),
\end{equation}
where $\text{erf}(\cdot)$ denotes the error function.

\textbf{Uniform Noise:} $e_i \sim \mathcal{U}(-\varepsilon, \varepsilon)$ independently.
Then, $e_1-e_2 \sim \text{Triangular}(-2\varepsilon, 2\varepsilon)$, i.e., 
\begin{equation} \label{eq:uniform_prob}
\mathbb{P}(|e_1-e_2| < 2\tau_j) =
\begin{cases}
1, &  \tau_j > \varepsilon, \\
\frac{2\tau_j}{\varepsilon}-\frac{\tau^2_j}{\varepsilon^2}  & \tau_j \leq \varepsilon.
\end{cases}
\end{equation}

\paragraph*{System Level Design}
The analytical expressions for various signal/noise distributions can be directly substituted into~\eqref{eq:success_prob_general} to compute theoretical success probabilities, enabling system designers to estimate performance without exhaustive simulations. These closed-form approximations guide parameter selection, including the number of robustness layers $K$, moduli pair $(\Gamma_1, \Gamma_2)$, and error threshold $m \sigma_j/2$ per layer. For instance, Rayleigh-distributed signal magnitudes concentrate probability mass near the origin, making lower robustness layers (small $j$) most critical for reconstruction.

However, the lower bound in Proposition~\ref{prop:success_rate} will \emph{become loose} for large $K$ due to non-uniform spacing of valid decoding regions. As shown in Fig.~\ref{fig:non_uniform}, slanted line gaps are non-uniform for small $j$ when $K\ge 3$, and larger gaps between adjacent lines reduce captured probability mass despite the minimum separation being lower bounded by $m \sigma_j$. This will be verified through simulations in Sec.~\ref{sec:simul}.

\section{Simulations}
\label{sec:simul}

In this section, we present simulation results to evaluate the effectiveness of the proposed moduli selection methods and to analyze the robustness performance of layered RCRT systems under different signal and noise models.

\begin{figure}[t]
    \centering
    \includegraphics[width=0.7\linewidth]{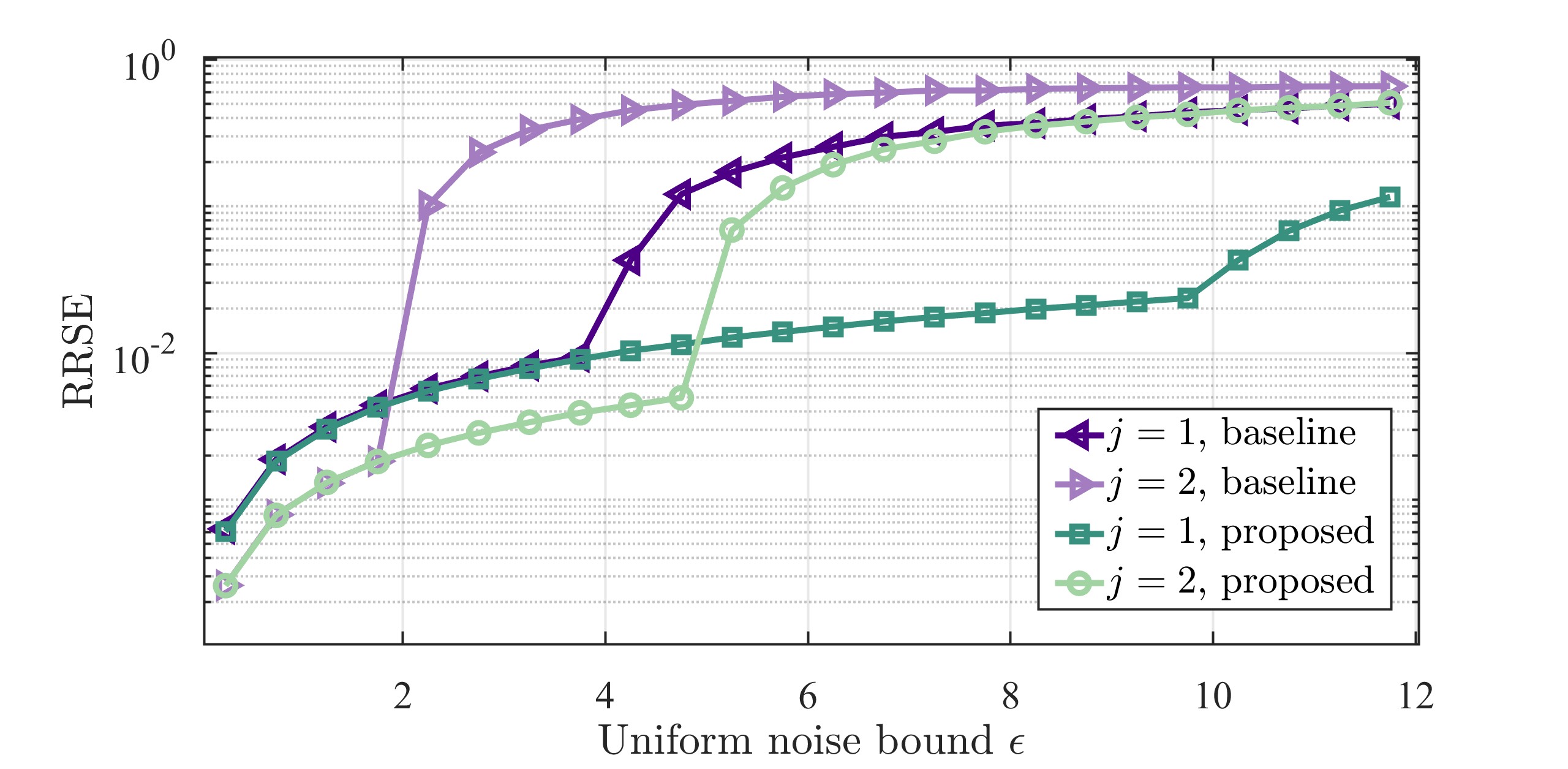}
\caption{RRSE vs \(\epsilon\) for two-moduli system with \(K = 1\), \(N_{\text{th}} = 680\), and \(m_{\max} = 136\), where remainder errors \(e \sim \mathcal{U}(-\epsilon, \epsilon)\), \(\epsilon \in [0, 12]\). 
\textbf{(a)} Baseline design from~\cite{xiao_towards_2017}: \((\Gamma_1, \Gamma_2, m) = (5, 17, 8)\), yielding \((P_1, \tau_1) = (285, 4)\), \((P_2, \tau_2) = (680, 2)\). 
\textbf{(b)} Proposed design: \((\Gamma_1, \Gamma_2, m) = (5, 7, 19\tfrac{3}{7})\), yielding \((P_1, \tau_1) = (291\tfrac{3}{7}, 9\tfrac{5}{7})\) and \((P_2, \tau_2) = (680, 4\tfrac{6}{7})\).}
    \label{fig:result1}
\end{figure}

\textbf{Simulation 1.} We evaluate the optimal moduli construction in Theorem~\ref{thm:L2L3Optimal} against randomly chosen moduli.
We fix \(N_{\text{th}} = 10^5\) and test two scenarios: \(L = 3\) with \(\rho \in\{ 100, 200\}\), and \(L = 4\) with \(\rho \in\{ 1000,2000\}\). For each case, we perform \(10^5\) Monte Carlo trials to estimate the success rate \(\beta_{\text{succ}}\) of random modulus selection in satisfying both the pairwise coprimality constraint~\eqref{eq:pairwise_coprime} and the dynamic range requirement~\eqref{eq:range_constraint}. For \(L = 3\), three integers are drawn uniformly at random from \([2, 100]\). For \(L = 4\), three odd integers and one even integer are sampled from the same range to ensure parity diversity.

Table~\ref{Table:Random} summarizes the results. Random selection exhibits low success rates and poor average robustness. While optimal bounds are occasionally achieved for \(L = 3\), the average \(\tau\) remains low. For \(L = 4\), the success rate is slightly higher due to the imposed parity diversity, but achieving the optimal bound becomes even more difficult, and the average \(\tau\) further decreases.

\begin{table}[t]
\centering
\caption{Comparison of Random and Optimal Moduli Selection under Different $L$ and $\rho$}
\renewcommand{\arraystretch}{1.5}
\begin{tabular}{cccccc}
\toprule
\toprule
$L$ &$\rho$ & Selection Strategy & $\beta_{\text{succ}}$ & $[\tau_{\min}, \tau_{\max}]$ & $\tau_{\text{avg}}$  \\
\midrule
\multirow{4}{*}{3} & \multirow{2}{*}{100} & Random & $30.07\%$ & $[5, 19.23]$ & $6.51$ \\
\cline{3-6}
& & Optimal & 1 & 19.23 & 19.23 \\
\cline{2-6}
& \multirow{2}{*}{200} & Random & $28.1\%$ & $[2.5, 7.35]$ & 3.17  \\
\cline{3-6}
& & Optimal & 1 & 7.35 & 7.35  \\
\midrule
\multirow{4}{*}{4} & \multirow{2}{*}{1000} & Random & $35.24\%$ & $[2.5,17.86]$ & 3.27  \\
\cline{3-6}
& & Optimal & 1 & 19.23 & 19.23  \\
\cline{2-6}
& \multirow{2}{*}{2000} & Random & $34.26\%$ & $[1.25, 7.35]$ & 1.62  \\
\cline{3-6}
& & Optimal & 1 & 8.33 & 8.33  \\
\bottomrule
\bottomrule
\end{tabular}\label{Table:Random}
\end{table}

\textbf{Simulation 2.} 
We revisit Example~1 from~\cite{xiao_towards_2017}, which uses moduli \((\Gamma_1, \Gamma_2) = (5, 17)\) and a common factor \(m = 8\), yielding a two-layer system with \((P_1, \tau_1) = (285, 4)\) and \((P_2, \tau_2) = (680, 2)\).
Under the same dynamic range \(N_{\text{th}} =\Gamma_1\Gamma_2m= 680\) and modulus bound \(m_{\max} =\Gamma_2m= 136\), our proposed design in Theorem~\ref{thm2:miniGamma}  produces \((\Gamma_1, \Gamma_2) = (5, 7)\) and \(m =\frac{136}{7}= 19\tfrac{3}{7}\), resulting in \((P_1, \tau_1) = (291\tfrac{3}{7}, 9\tfrac{5}{7}))\) and \((P_2, \tau_2) = (680, 4\tfrac{6}{7})\). Relative to the baseline from~\cite{xiao_towards_2017}, the optimized moduli nearly double the tolerance at both layers, with negligible change in \(P_1\) and no loss in \(P_2\).

To empirically verify the improvement, we carry out Monte Carlo simulations under both configurations.
Both settings adopt the same two-level decoding: Algorithm~1 from~\cite{xiao_towards_2017} is used at the first layer, and the closed-form RCRT in~\cite{CRT} at the second. In each trial, the ground-truth \(x\) is uniformly drawn from \([1, P_j)\), with remainder noise sampled from \([{-}\epsilon, \epsilon]\), \(\epsilon \in (0,12]\). The Root Relative Squared Error (RRSE) over \(R = 10^4\) Monte Carlo trials is given by
\begin{equation}
\text{RRSE} = \sqrt{ \frac{ \sum_{i=1}^{R} (x_i - \tilde{x}_i)^2 }{ \sum_{i=1}^{R} x_i^2 } },
\label{eq:rrse_mc}
\end{equation}
where \(x_i\) and \(\tilde{x}_i\) denote the true and reconstructed values in the \(i\)th trial.
As shown in Fig.~\ref{fig:result1}, the proposed moduli improve robustness at both levels, doubling the error tolerance with similar \(P_\ell\) ($\ell=1,2$). This confirms the practical advantage of our design under fixed constraints.

\textbf{Simulation 3.} We study how $K$ affects reconstruction under Rayleigh-distributed inputs: $x \sim \mathrm{Rayleigh}(360)$, with dynamic range $N_{\text{th}} = 1800$ (outlier probability approximately $3.73 \times 10^{-6}$). The modulus bound is $m_{\max} = 120$, yielding $\rho = N_{\text{th}}/m_{\max} = 15$, and $K^* = 5$ via~\eqref{eq:Kstar}. Moduli pairs $(\Gamma_2, \Gamma_1)$ are selected using Theorem~\ref{thm:L2L3Optimal} for $K = 0$ and Theorem~\ref{thm2:miniGamma} for $K \in \{1, 2, 3, 4, 10\}$.

The remainder noise $e_i$ ($i=1,2$) is drawn either from a zero-mean Gaussian distribution \(e_i \sim \mathcal{N}(0, \varsigma^2)\) or a uniform distribution \(e_i \sim \mathcal{U}(-\epsilon, \epsilon)\), where \(\varsigma, \epsilon \in [0,6]\) with a step size of $0.5$. For reconstruction, the RCRT from~\cite{CRT} is used at the final layer (\(j = K+1\)). For \(K \ge 1\), layers \(j=1\) and \(2 \le j \le K\) are decoded via Algorithms~1 and~2 of~\cite{xiao_towards_2017}, respectively. Each success rate \(\eta_{\text{succ}}\) is estimated from \(10^5\) Monte Carlo trials.

\begin{figure*}[t]
    \centering
    \includegraphics[width=0.8\linewidth]{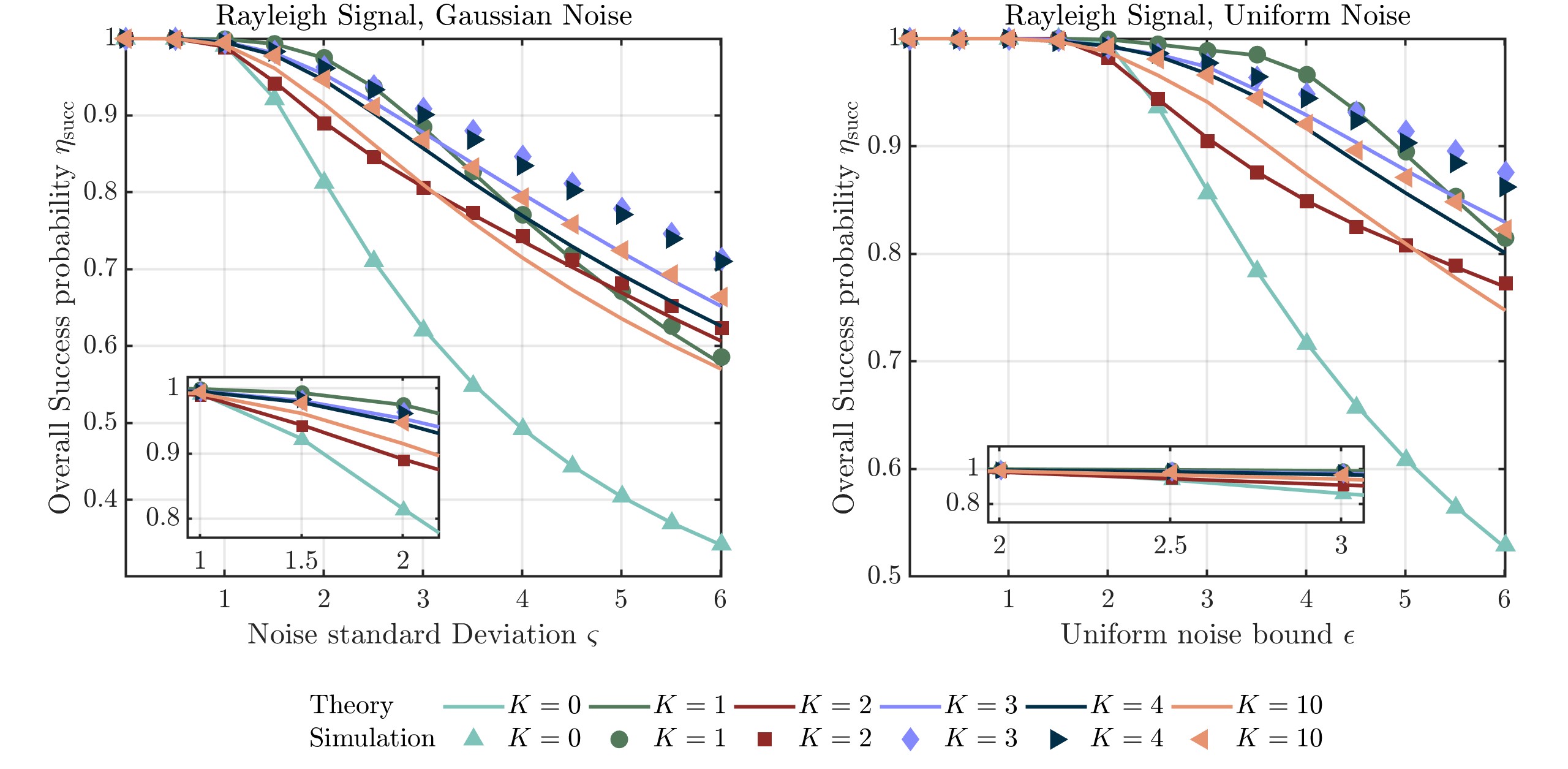}
\caption{Overall success probability \(\eta_{\text{succ}}\) vs noise intensity under Rayleigh-distributed signals with scale \(\beta = 360\). 
Left: Gaussian noise \(e_i \sim \mathcal{N}(0, \varsigma^2)\), \(\varsigma \in [0, 6]\), step size 0.5. 
Right: Uniform noise \(e_i \sim \mathcal{U}(-\epsilon, \epsilon)\), \(\epsilon \in [0, 6]\), step size 0.5. 
Solid lines show theoretical predictions; markers indicate simulated results for recursion depths \(K \in \{0, 1, 2, 3, 4, 10\}\).}
\label{fig:RayleighNoiseComparison}
    \label{fig:RayonK}
\end{figure*}

Fig.~\ref{fig:RayonK} presents both theoretical predictions (solid lines) and empirical results (markers) for overall success probability as noise intensity increases. The left plot corresponds to Gaussian noise (parameterized by standard deviation \(\varsigma\)), while the right shows uniform noise (with amplitude bound \(\epsilon\)). The insets zoom in on the high-success region. From this figure, one can observe:
\begin{itemize}
\item For small noise (e.g., \(\varsigma \lesssim 1\), \(\epsilon \lesssim 1.5\)), all configurations achieve near-perfect success.
\item As noise increases, adding robust layers ($K\geq 1$) exhibit slower degradation and outperform the baseline \(K=0\) by a wide margin.
\item For \(K \le 2\), theoretical bounds closely match the simulation results. For larger \(K\), the empirical performance increasingly exceeds the bound, highlighting its looseness as $K$ increases, a behavior predicted by our earlier analysis.
\item The most substantial gains occur for \(K = 1\) or \(2\). Beyond \(K = 3\), improvements are not obvious due to the geometric narrowing of admissible plateaux (\(P_{j+2}/P_j \to \varphi^2\)).
\end{itemize}

\textbf{Analogy to $\mu$-law and A-law companding.} 
Classical $\mu$-law and A-law quantizers apply a logarithmic companding curve so that small amplitudes receive finer resolution and large amplitudes coarser resolution.  The layered RCRT architecture exhibits a similar non-uniform treatment in the remainder domain.  Its staircase error-tolerance curve \(T(x)\) takes higher values for small \(|x|\) and progressively smaller values as it \(|x|\) increases, while the dynamic-range intervals \([P_{j-1},P_j)\) behave conversely: narrow near the origin and wider at large magnitude.  Hence, low-amplitude samples enjoy large error margins but occupy short ranges, whereas high-amplitude samples see tighter margins over longer ranges.
This built-in “unequal error protection’’ is well matched to sources such as speech, audio, or communication signals, whose probability density of signal magnitude is high near zero. Choosing one or two robust layers (\(K = 1\) or \(2\)) therefore protects the portion of the signal that occurs most frequently, much as the knee of a companding curve preserves small-signal fidelity.

\textbf{Design Implications}:   
The proposed layered RCRT framework supports flexible trade-offs between range and robustness, offering the following practical insights:

\begin{itemize}
    \item \textbf{Signal statistics:} For signals concentrated near the origin, e.g., Rayleigh or folded Gaussian distributions, adding one or two robust layers (\(K=1\) or \(2\)) significantly improves decoding success rates, since a high proportion of signal magnitudes fall within the inner plateaux. 
    
    \item \textbf{Resource-constrained systems:} In resource-constrained applications, e.g., with limited buffer size or processing power, small $K$ (\(K \leq 2\)) offers improved robustness without significant complexity overhead.
    
    \item \textbf{Multi-resolution applications:} Layered RCRT can support multi-user or multi-resolution reconstruction, where each user operates at a layer tuned to their dynamic range requirement. Moderate values of \(K\) (e.g., $K=3,4$) are especially useful in such scenarios.
    
    \item \textbf{Limited returns for large \(K\):} Adding many layers (\(K > K^*\)) provides only marginal improvement in success probability at low signal-to-noise ratio (SNR), as the robust range contracts and the dominant failure events are not further mitigated. As shown in Fig.~\ref{fig:RayleighNoiseComparison}, performance curves flatten beyond \(K=3\), indicating that deep layering may not be cost-effective for most use cases.
\end{itemize}

\section{Conclusion and Future Work}  
\label{sec:conclusion}  
This paper presents a unified design framework for moduli selection in RCRT decoding algorithms~\cite{CRT,xiao_towards_2017}. For the basic RCRT system with a single layer, we derive closed-form optimal moduli for \(L = 2, 3,\) and \(4\) that maximize error tolerance. For multi-layer systems, we propose a Fibonacci-inspired construction that ensures exactly \(K\) robust decoding layers in addition to the full CRT layer. We further analyze how the number of robust layers \(K\) affects both the dynamic range and the error tolerance, and derive closed-form success probabilities under typical signal and noise models, offering system-level insights for practical design. All theoretical developments are supported by examples and simulations, confirming the validity and practical potential of the proposed methods. Simulation results demonstrate that adding one or two robust layers (i.e., \(K = 1, 2\)) offers substantial robustness gains with minimal compromise in full CRT range, which makes the method especially well-suited to practical applications.

Future work may include tighter probabilistic analysis of success rates for \(K \geq 3\), generalization to multi-moduli layered systems (\(L > 2\)), and alternative optimization objectives beyond error tolerance at the full CRT layer. To further enhance robustness, one may also incorporate distributional uncertainty, adaptive modeling, or Bayesian priors in reconstruction, depending on the target applications. 

\section*{Appendix A\\ Proof of Theorem~\ref{thm:L2L3Optimal}}

\begin{IEEEproof}

\textbf{Case \( L = 2 \):}
By~\eqref{eq:rho_def}, we must have \( \Gamma_1 \geq \lceil \rho \rceil \). To minimize \( \Gamma_2 \), we select \( \Gamma_2 = \Gamma_1 + 1 \).  
From Lemma~\ref{lem:coprime_properties}, they are coprime, which means this selection is optimal.

\vspace{1ex}


\textbf{Case $L=3$}: \emph{Trivial regime $1<\rho\le 6$):}
The smallest sorted coprime pair is $(\Gamma_1,\Gamma_2)=(2,3)$, which
meets the range constraint since $\Gamma_1\Gamma_2=6\ge \rho$.
The minimal admissible $\Gamma_3>\Gamma_2$ coprime to both $2$ and $3$ is $5$
($4$ is excluded as it shares a factor with $2$).
Hence $(\Gamma_1,\Gamma_2,\Gamma_3)=(2,3,5)$ is feasible and minimizes $\Gamma_3$.

\textit{General regime \(\rho>6\):}
Since $\Gamma_1\le \Gamma_2-1$ and $\rho$ is defined in \eqref{eq:rho_def}, we have
$\Gamma_2(\Gamma_2-1)\ge \Gamma_2\Gamma_1\ge \rho$. Let $b$ be the smallest
integer (with $b\ge 3$) such that $b(b-1)\ge \rho$ (cf.~\eqref{eq:defb}).
Then necessarily
\[
\Gamma_2\ \ge\ b,\qquad \Gamma_3\ \ge\ b+1.
\]

\emph{Even $b$.}
The triple $\{\Gamma_1,\Gamma_2,\Gamma_3\}=\{b-1,b,b+1\}$ achieves the lower
bound $\Gamma_3=b+1$, satisfies $\Gamma_1\Gamma_2=(b-1)b\ge \rho$, and is
pairwise coprime by Lemma~\ref{lem:coprime_properties}. Hence it is feasible
and has minimal $\Gamma_3$.

\emph{Odd $b$.}
First try $(\Gamma_2,\Gamma_3)=(b,b+1)$ and ask whether there exists
$\Gamma_1<b$ with $\Gamma_1\Gamma_2\ge \rho$ and pairwise coprimality.

\smallskip
\underline{$(b+1)\not\equiv 0\pmod{3}$.}
Then $\gcd(b-2,b+1)=\gcd(3,b+1)=1$, and by
Lemma~\ref{lem:coprime_properties} also $\gcd(b-2,b)=\gcd(b,b+1)=1$.
If $b(b-2)\ge \rho$, choosing
\[
(\Gamma_1,\Gamma_2,\Gamma_3)=(b-2,b,b+1)
\]
meets the range constraint with pairwise coprimality and achieves
$\Gamma_3=b+1$, which is minimal.

\smallskip
\underline{$(b+1)\equiv 0\pmod{3}$.}
Now $\gcd(b-1,b+1)=2$ and $\gcd(b-2,b+1)=3$, so neither $\Gamma_1=b-1$ nor
$\Gamma_1=b-2$ is coprime to $\Gamma_3=b+1$. Moreover, for any $\beta\ge 3$,
\[
b(b-\beta)\ \le\ b(b-3)\ <\ (b-1)(b-2)\ <\ \rho,
\]
so any $\Gamma_1\le b-3$ violates the range constraint. Thus no
$\Gamma_1<b$ works when $\Gamma_3=b+1$, and therefore $\Gamma_3\ge b+2$ is
necessary. One checks that
\[
(\Gamma_1,\Gamma_2,\Gamma_3)=(b,b+1,b+2)
\]
is pairwise coprime and satisfies $\Gamma_1\Gamma_2=b(b+1)\ge \rho$, so it is
feasible and has the smallest admissible $\Gamma_3$ in this subcase.

\smallskip
Finally, if $b(b-2)<\rho$ (still with $b$ odd), the choice
$(\Gamma_1,\Gamma_2)=(b-2,b)$ cannot meet the range constraint, and the above
argument again forces $\Gamma_3\ge b+2$; the triple $(b,b+1,b+2)$ is feasible
and minimal.

\vspace{1ex}
\textbf{Case $L = 4$:}
\noindent\textit{Trivial regime \(1<\rho\le 30\).}
The smallest pairwise–coprime triple is \((2,3,5)\); thus we set
\((\Gamma_1,\Gamma_2,\Gamma_3)=(2,3,5)\), which satisfies the range constraint since
\(\Gamma_1\Gamma_2\Gamma_3=30\ge \rho\).
To minimize \(\Gamma_4\) subject to ordering and coprimality with \((2,3,5)\),
the smallest admissible value \(>\!5\) is \(7\) (as \(6\) shares factors with \(2,3\)).
Therefore \((\Gamma_1,\Gamma_2,\Gamma_3,\Gamma_4)=(2,3,5,7)\) is feasible and minimizes \(\Gamma_4\)
for all \(1<\rho\le 30\).

\textit{General regime \(\rho>30\):}
We verify, for each row of Table~\ref{tab:opt4}, that (\emph{i}) the product constraint holds,
(\emph{ii}) pairwise coprimality holds (via Lemma~\ref{lem:coprime_properties} and the stated congruence classes),
and (iii) \(\Gamma_4\) is minimal by excluding any smaller admissible value.
Let \(b\) be defined by \eqref{eq:defb_L4}; then necessarily \(\Gamma_3\ge b\) and \(\Gamma_4\ge b+1\).
When \(\Gamma_4=b+1\) is feasible, optimality is immediate; when \(\Gamma_4\in\{b+2,b+3\}\),
we show any smaller choice violates (\emph{i}) or (\emph{ii}).

\subsection*{Case A: odd \(b\)}
\textbf{Rows A1–A2.}
From \eqref{eq:defb_L4},
\begin{equation}\label{eq:bL4prop}
(b-1)(b-2)(b-3)\ \le\ \rho\ <\ b(b-1)(b-2).
\end{equation}
Both quartets \(\{b-2,b-1,b,b+2\}\) (A1) and \(\{b-2,b,b+1,b+2\}\) (A2) satisfy the product bound.
For coprimality: \(\{b-2,b-1,b\}\) and \(\{b-2,b,b+2\}\) are pairwise coprime (Lemma~\ref{lem:coprime_properties}).
Moreover, \(\gcd(b-1,b+2)=1\) when \(b\not\equiv1\pmod{3}\) (A1), and the analogous checks hold for A2.
For optimality, suppose a quartet with \(\Gamma_4=b+1\) exists. Since \(b+1\) is even,
the remaining three moduli must be odd. Note that the three largest \(<b+1\) are \(b,\,b-2,\,b-4\).
But then
\[
(b-4)(b-2)b\ <\ (b-1)(b-2)(b-3)\ \le\ \rho,
\]
contradicting the product constraint. Hence \(\Gamma_4\ge b+2\) and the listed quartets are optimal.

\subsection*{Case B: Even $b$, $(b-3)(b-1)b > \rho$}

\textbf{Row B1.} ($b \bmod 6 \in \{2,4\}$)

The product requirement is satisfied by assumption. Pairwise coprimality holds for $\{b-3,b-1,b+1\}$ and $\{b-1,b,b+1\}$, and $\gcd(b,b-3)=1$ since $b\bmod 3\neq 0$. Optimality follows from $\Gamma_4\geq b+1$.

\textbf{Row B2.} ($b \bmod 6 = 0$, $b \bmod 5 \neq 3$)

The product constraint holds for $\{b-3,b-1,b+1\}$ by assumption. The sets $\{b-3,b-1,b+1\}$ and $\{b+1,b+2\}$ are pairwise coprime. In addition, $\gcd(b-1,b+2)=1$ (since $b\bmod 3\neq 1$) and $\gcd(b+2,b-3)=1$ (since $b\bmod 5\neq 3$). Hence, coprimality holds.

Assume that $\Gamma_3=b$ and $\Gamma_4=b+1$. Since $b$ is even, the remaining two moduli must be odd. However, $\gcd(b,b-3)=3$ excludes $b-3$. Thus, the largest available odd numbers are $b-1$ and $b-5$. Then:
\[
b(b-1)(b-5) < (b-1)(b-2)(b-3) < \rho,
\]
violating the product requirement. Hence, $\Gamma_3\geq b+1$ and $\Gamma_4\geq b+2$, proving optimality.

\textbf{Row B3.} ($b \bmod 6 = 0$, $b \bmod 5 = 3$)

From the proof of Row B2, when $b \bmod 6 = 0$: (\emph{i}) $\Gamma_3\geq b+1$, $\Gamma_4\geq b+2$, and (\emph{ii}) $\gcd(b,b-3)=3$, so $b$ and $b-3$ cannot coexist. Additionally, $\gcd(b+2,b-3)=5$, so $b-3$ and $b+2$ cannot coexist.

\paragraph*{Row B3.1.} ($b \bmod 6 = 0$, $b \bmod 5 = 3$, $(b-5)(b-1)(b+1) > \rho$ and $b+2 \bmod 7 \neq 0$)

The quartet $\{b-5,b-1,b+1,b+2\}$ satisfies the product constraint. By Lemma~\ref{lem:coprime_properties}, $\{b-1,b+1,b+2\}$ are pairwise coprime. We also have $\gcd(b-5,b-1)=\gcd(4,b-1)=1$, $\gcd(b-5,b+1)=1$ (since $b\bmod 6=0$), and $\gcd(b-5,b+2)=1$ (since $b+2\bmod 7\neq 0$). Optimality follows from the result of B2.

\paragraph*{Row B3.2.}

The quartet $\{b-1,b+1,b+2,b+3\}$ satisfies $(b-1)(b+1)(b+2)>\rho$. Pairwise coprimality is satisfied for $\{b-1,b+1,b+3\}$ and $\{b+1,b+2,b+3\}$ from Lemma~\ref{lem:coprime_properties}. Also, $\gcd(b+2,b-1)=1$ since $b\bmod 6=0$. Thus, all pairs are coprime.

To prove optimality, assume $\Gamma_4=b+2$. Then all other moduli must be odd. We cannot select $b-3$ since $\gcd(b+2,b-3)=5$). In addition, either $(b-5)(b-1)(b+1)\leq\rho$ or $\gcd(b+2,b-5)=7$ excludes $b-5$. The remaining choice is $\{b+1, b-1, b-7\}$, but:
\[
(b-7)(b-1)(b+1) < (b-5)(b-1)(b+1) \leq \rho,
\]
which violates the product constraint. Hence, $\Gamma_4\geq b+3$, and the construction is optimal.

\subsection*{Case C – Even \texorpdfstring{$b$}{b} with \texorpdfstring{$(b-3)(b-1)b < \rho \le (b-3)(b-1)(b+1)$}{product between bounds}}

\vspace{1ex}
\textbf{Row C1} ($b\bmod3\neq1$, $b\bmod5\neq3$).

The quartet $\{b-3,b-1,b+1,b+2\}$ satisfies the product bound because
$(b-3)(b-1)(b+1)>\rho$. Pairwise coprimality follows from Lemma~\ref{lem:coprime_properties}: $\gcd(b-3,b+2)=1$ since $b\bmod5\neq3$, and $\gcd(b-1,b+2)=1$ since $b\bmod3\neq1$.
To prove optimality, assume $\Gamma_4=b+1$.  Then $\Gamma_3= b$ (even) 
and the two remaining moduli must be odd.  The largest
such pair is $\{b-1,b-3\}$. As
$(b-3)(b-1)b\le\rho$, the product condition is violated.
Any smaller odd choice further reduces the product, so $\Gamma_4=b+1$
is impossible; hence $\Gamma_4\ge b+2$, and the selected quartet is
optimal.

\vspace{1ex}
\textbf{Row C2} ($b\bmod3=1$, $b\bmod5\neq3$).  

The proof of the product and pairwise property is straightforward.

To show optimality, the proof of C1 already indicates that $\Gamma_4\ge b+2$. Assume for contradiction that $\Gamma_4 = b+2$ and $\Gamma_3 = b+1$. The remaining two moduli must be odd and less than $b+1$.  
However, $b-1$ is excluded because $\gcd(b+2,b-1)=\gcd(3, b-3)=3$ (as $b\bmod3=1$).  
Thus, the largest valid odd pair is $\{b-5, b-3\}$. Again, as  
\[
(b-5)(b-3)(b+1) < (b-3)(b-1)b < \rho,
\]
the product constraint is not met. Hence, $\Gamma_4=b+2$ is infeasible and the chosen quartet with $\Gamma_4 = b+3$ is optimal.

\vspace{1ex}
\textbf{Row C3} ($b\bmod5=3$).

\textit{Product:} By definition of $b$, the product constraint is satisfied as $(b-1)(b+1)(b+2)> (b+1)b(b-1)>\rho$.

\textit{Coprimality:} The set depends on $b\bmod3$.  
If $b\bmod3=1$, then $\gcd(b+2,b-1)=\gcd(3, b-3)=3$, so $b-1$ and $b+2$ cannot be used together.  
We instead select $\{b-1,b,b+1,b+3\}$, which is pairwise coprime by Lemma~\ref{lem:coprime_properties}.  
Otherwise, use $\{b-1,b+1,b+2,b+3\}$, which is also pairwise coprime (proof as in B3.2).

\textit{Optimality:} As mentioned before, the proof of C1 already establishes that $\Gamma_4\ge b+2$. Suppose, for contradiction, that $\Gamma_4 = b+2$.  
As $b\bmod 5 = 3$, we must exclude $b-3$ since $\gcd(b+2,b-3) = \gcd(b+2,5) = 5$.  
Thus, the largest three odd integers below $b+2$ are $b+1$, $b-1$, and $b-5$, whose product is
\[
(b-5)(b-1)(b+1) < (b-3)(b-1)b < \rho,
\]
contradicting the product constraint. Hence, $\Gamma_4 \ge b+3$, which confirms optimality of the selected quartet.

\vspace{1ex}
\subsection*{Case D — even $b$ with $(b+1)(b-1)(b-3)<\rho$}


\textit{Product:} As  $b(b-1)(b-2)\ge \rho$, it is trivial to prove that both  selected quartet $\{b-1,b,b+1,b+3\}$ 
and $\{b-1,b+1,b+2,b+3\}$  satisfy the product requirement.

\textit{Coprimality:} For Row D1, the subsets $\{b-1,b+1,b+3\}$ and $\{b-1,b,b+1\}$ are pairwise coprime by Lemma~\ref{lem:coprime_properties}. Since $b \bmod 3 \neq 0$, we have $\gcd(b,b+3)=1$, guaranteeing the full quartet's pairwise coprimality.

For Row D2, the proof is analogous. The subsets $\{b-1,b+1,b+3\}$ and $\{b+1,b+2,b+3\}$ are pairwise coprime by Lemma~\ref{lem:coprime_properties}. Moreover, with $b$ even and $b \bmod 3 = 0$, we obtain $\gcd(b+2,b-1) = \gcd(b-1,3) = 1$, thus ensuring the full quartet's pairwise coprimality.

\textit{Optimality:}  We demonstrate that the minimal pair $(\Gamma_4,\Gamma_3)=(b+1,b)$ cannot be selected. Under this choice, the remaining moduli must be odd since $b$ is even, yielding the largest available odd values $(\Gamma_2,\Gamma_1)=(b-1,b-3)$. However, the assumption $(b-3)(b-1)b<(b-3)(b+1)<\rho$ violates the product constraints.

We proceed to establish that $\Gamma_4 = b+2$ is also infeasible. This selection leaves $b+1$, $b-1$, and $b-3$ as the three largest available odd integers. The corresponding assumption $(b-3)(b-1)(b+1) < \rho$ again violates the product constraint. Therefore, $\Gamma_4 \ge b+3$, and since the proposed quartet achieves this lower bound, optimality is established.

\end{IEEEproof}

\section*{Appendix B: Proof of Theorem~\ref{thm2:miniGamma}}
\subsection*{Part (i): Remainder Chain and Layered Dynamic Range}

\noindent\textbf{1) Coprimality.}
Clearly $\Gamma_1(d)\ge\rho$ by construction. Let $g=\gcd(\Gamma_1(d),\Gamma_2(d))$. Since $g$ divides both numbers, it divides their
difference:
\[
\Gamma_2(d)-\Gamma_1(d)=F_{d,K+1}\quad\Rightarrow\quad g\,\mid\,F_{d,K+1}.
\]
Also $g\,\mid\,\Gamma_1(d)$ implies
\[
g\,\mid\,\big(\Gamma_1(d)-\zeta_d F_{d,K+1}\big)=F_{d,K}.
\]
Hence $g$ divides both $F_{d,K}$ and $F_{d,K+1}$. By the Euclidean algorithm,
$\gcd(F_{d,K},F_{d,K+1})=\gcd(F_{d,0},F_{d,1})=\gcd(d,1)=1$, so $g=1$.

\smallskip
\noindent\textbf{2) Exact remainder chain.}:
$\sigma_{-1}=\Gamma_2(d)$ and $\sigma_{0}=\Gamma_1(d)$. A simple calculation gives
\(
\sigma_1=\Gamma_2(d)-\Gamma_1(d)=F_{d,K+1}.
\)
Assume inductively
\(
\sigma_{j}=F_{d,K+2-j}
,\;
\sigma_{j-1}=F_{d,K+3-j}
\)
for some $j\ge1$.
Using $F_{d,k+2}=F_{d,k+1}+F_{d,k}$,
\[
\sigma_{j+1}=\sigma_{j-1}\bmod\sigma_{j}
           =\sigma_{j-1}-\sigma_{j}=F_{d,K+1-j},
\]
so by induction
\(
\sigma_j=F_{d,K+2-j}\;(1\le j\le K+1)
\)
and the algorithm terminates after $K{+}1$ steps at
$\sigma_{K+1}=F_{d,1}=1$.

\noindent\textbf{3) Layered dynamic range.}
Let $P_j=m\,N_j$. Proving \eqref{eq:Pj_compact} is equivalent to showing that, for $1\le j\le K$,
\begin{equation}\label{eq:Nj_proposed}
N_j=
\begin{cases}
\Gamma_1(d)\,F_{\zeta_d,\,2j_0}, & j=2j_0-1,\\[4pt]
\Gamma_2(d)\,F_{\zeta_d-1,\,2j_0+1}, & j=2j_0,
\end{cases}
\end{equation}
where $j_0\in\mathbb{Z}_{\ge1}$. We proceed in three steps.

\smallskip
\emph{(a) Two auxiliary identities.}
We first prove that
\begin{align}
&\Gamma_2(d)\,F_{\zeta_d-1,\,j}-\Gamma_1(d)\,F_{\zeta_d,\,j}=\sigma_j,\quad  j\text{ odd}, \label{eq:odd_aux}\\
&\Gamma_1(d)\,F_{\zeta_d,\,j}-\Gamma_2(d)\,F_{\zeta_d-1,\,j}=\sigma_j,\quad  j\text{ even}. \label{eq:even_aux}
\end{align}
Let us consider the case of odd $j=2j_0-1$. For $j=1$ (i.e., $j_0=1$), $F_{\zeta_d-1,1}=F_{\zeta_d,1}=1$, hence
\[\Gamma_2(d)-\Gamma_1(d)=F_{d,K+1}=\sigma_1,\] proving \eqref{eq:odd_aux}.
For $j=2j_0-1>1$, write
$F_{\zeta_d-1,j}=F_{1,j-1}+(\zeta_d-1)F_{1,j-2}
=F_{\zeta_d,j}-F_{1,j-2}$,
and use the equality of $\Gamma_2(d)-\Gamma_1(d)=F_{d,K+1}$ from \eqref{eq:gamma12-d} to obtain
\[
\Gamma_2(d)F_{\zeta_d-1,j}-\Gamma_1(d)F_{\zeta_d,j}
=F_{d,K+1}F_{\zeta_d,j}-\Gamma_2(d)F_{1,j-2}.
\]
With $F_{\zeta_d,k}=F_{1,k-1}+\zeta_d F_{1,k-2}$ and
$\Gamma_2(d)=(\zeta_d+1)F_{d,K+1}+F_{d,K}$, a short cancellation yields
\[
\Gamma_2(d)F_{\zeta_d-1,j}-\Gamma_1(d)F_{\zeta_d,j}
=-(F_{d,K}F_{1,j-2}-F_{d,K+1}F_{1,j-3}).
\]
Applying the mixed d\textquotesingle Ocagne identity (Lemma~\ref{lemma:FibLikeProperties}) with $s=K$ and $t=j-2$ gives
\[-(F_{d,K}F_{1,j-2}-F_{d,K+1}F_{1,j-3})=(-1)^{j-1}F_{d,K+2-j}=\sigma_j\]
for $j=2j_0-1$, proving \eqref{eq:odd_aux}.
The even case \eqref{eq:even_aux} follows by the same argument with roles interchanged.

\smallskip
\emph{(b) Base layers $j=1,2$.}
From \eqref{eq:sigmaexp_xiao} and \eqref{eq:gamma12-d}, $\sigma_0=\Gamma_1(d)$ and
$\sigma_1=\Gamma_2(d)\bmod \Gamma_1(d)=F_{d,K+1}$, hence
$\big\lfloor\sigma_0/\sigma_1\big\rfloor=\zeta_d$.
Using \eqref{eq:Nj_xiao_alt} with $N_{-1}=\Gamma_1(d)$ and $N_0=\Gamma_2(d)$,
\[
\begin{split}
N_1
&= N_{-1}+\Big\lfloor\frac{\sigma_0}{\sigma_1}\Big\rfloor\,(N_0-\sigma_1)\\
&= \Gamma_1(d)+\zeta_d\big(\Gamma_2(d)-F_{d,K+1}\big)\\&= \Gamma_1(d)+\zeta_d\Gamma_1(d)
= \Gamma_1(d)F_{\zeta_d,2},
\end{split}
\]
so \eqref{eq:Nj_proposed} holds for $j=1$.
For $j=2$, since $\big\lfloor\sigma_1/\sigma_2\big\rfloor
=\big\lfloor F_{d,K+1}/F_{d,K}\big\rfloor=1$ and \eqref{eq:even_aux} with $j=2$
gives $\sigma_2=\Gamma_1(d)F_{\zeta_d,2}-\Gamma_2(d)F_{\zeta_d-1,2}$,
we obtain
\[
\begin{aligned}
N_2&=N_0+\Big\lfloor\tfrac{\sigma_1}{\sigma_2}\Big\rfloor\,(N_1-\sigma_2)
=\Gamma_2(d)+\Gamma_1(d)F_{\zeta_d,2}-\sigma_2\\
&=\Gamma_2(d)+\Gamma_2(d)F_{\zeta_d-1,2}
=\Gamma_2(d)F_{\zeta_d-1,3},
\end{aligned}
\]
verifying \eqref{eq:Nj_proposed} for $j=2$.

\smallskip
\emph{(c) Induction.}
Assume that \eqref{eq:Nj_proposed} holds for $(2j_0-1,2j_0)$ with $1\le j_0<\lceil K/2\rceil$,
i.e.,
$N_{2j_0-1}=\Gamma_1(d)F_{\zeta_d,\,2j_0}$ and
$N_{2j_0}=\Gamma_2(d)F_{\zeta_d-1,\,2j_0+1}$.
For $2\le j\le K$, the Euclidean remainders satisfy
$\big\lfloor\sigma_{j-1}/\sigma_j\big\rfloor
=\big\lfloor F_{d,K+3-j}/F_{d,K+2-j}\big\rfloor=1$.
Then
\[
\begin{aligned}
N_{2j_0+1}
&=N_{2j_0-1}
+\Big(N_{2j_0}-\sigma_{2j_0+1}\Big)\\
&=\Gamma_1(d)F_{\zeta_d,\,2j_0}
+\Big(\Gamma_2(d)F_{\zeta_d-1,\,2j_0+1}-\sigma_{2j_0+1}\Big)\\
&=\Gamma_1(d)F_{\zeta_d,\,2j_0}
+\Gamma_1(d)F_{\zeta_d,\,2j_0+1}
=\Gamma_1(d)F_{\zeta_d,\,2j_0+2},
\end{aligned}
\]
where we used \eqref{eq:odd_aux} with $j=2j_0+1$ in the third line.
Thus \eqref{eq:Nj_proposed} holds for $j=2j_0+1$.
The even step is analogous: using \eqref{eq:even_aux} with $j=2j_0+2$ gives
\[
\begin{aligned}
N_{2j_0+2}
&=N_{2j_0}
+\Big(N_{2j_0+1}-\sigma_{2j_0+2}\Big)\\
&=\Gamma_2(d)F_{\zeta_d-1,\,2j_0+1}
+\Big(\Gamma_1(d)F_{\zeta_d,\,2j_0+2}-\sigma_{2j_0+2}\Big)\\
&=\Gamma_2(d)F_{\zeta_d-1,\,2j_0+1}
+\Gamma_2(d)F_{\zeta_d-1,\,2j_0+2}
=\Gamma_2(d)F_{\zeta_d-1,\,2j_0+3}.
\end{aligned}
\]
By induction, \eqref{eq:Nj_proposed} holds for all $1\le j\le K$, and hence
$P_j=m\,N_j$ equals \eqref{eq:Pj_compact}.

\subsection*{Part (ii): Bounded Search for Optimal $d$}
\medskip
For $d=1$,
\(
\Gamma_2(1)=\bigl(\zeta_{1}+1\bigr)F_{1,K+1}+F_{1,K}.
\)
Recall that for each \( d>1 \), the construction yields $\Gamma_2(d)=\Gamma_1(d) + F_{d,K+1} $.
To ensure feasibility, we require \( \Gamma_1(d) \ge \rho \), which implies that
$\Gamma_2 (d) \geq \rho  + F_{d,K+1}$. 
By definition of $\zeta_1$, we have 
\[\rho>(\zeta_1-1)F_{1,K+1}+F_{1,K},\]
 which implies that 
 \[\Gamma_2 (d) > (\zeta_1-1)F_{1,K+1}+F_{1,K}+F_{d,K+1} \]
 To guarantee improvement over $\Gamma_2(1)$, i.e., $\Gamma_2(d)<\Gamma_2(1)$, we further require
\[
\Gamma_2(d) <(\zeta_1+1) F_{1,K+1} + F_{1,K},
\]
which produces 
\[
(\zeta_1-1)F_{1,K+1}+F_{1,K}+F_{d,K+1}<(\zeta_1+1) F_{1,K+1} + F_{1,K},  \]
and can be simplified to $F_{d,K+1}<2F_{1,K+1}$. Substituting \( F_{d,K+1} = F_{1,K} + d F_{1,K-1} \) from Lemma~\ref{lemma:FibLikeProperties} along with the recursion $F_{1,K+1}=F_{1,K}+F_{1,K-1}$, we know that $d$ needs to satisfy the following constraint
\begin{equation}\label{eq:d_2_range}
    (d-2) F_{1,K-1} < F_{1,K}.
\end{equation}
When $K=1$, as $F_{1,0}=1$ and $F_{1,1}=1$, we have $d<3$. When $K=2$, as $F_{1,2}=2$,
we get $d<4$, which indicates $d\le 3$. When $K\ge 3$, 
as $F_{1,K}<2F_{1,K-1}$, we arrive at
$(d-2) F_{1,K-1} <2F_{1,K-1}, 
$
implying $d<4$, i.e., $d\le 3$. Summarising, we only need to check $1\le d \le 3$.

\section*{Appendix C: Proof of Corollary~\ref{cor:explicit_opt_pairs}}
\label{app:pf_cor_explicit}


\begin{IEEEproof}
By Theorem~\ref{thm2:miniGamma}, the construction guarantees
$\Gamma_1\ge\lceil\rho\rceil$ and $\gcd(\Gamma_1,\Gamma_2)=1$. It therefore
suffices to show that $\Gamma_2$ is minimal under these constraints.

\medskip
\noindent\emph{A necessary gap at the first step.}
Let $\sigma_1=\Gamma_2\bmod \Gamma_1=\Gamma_2-\Gamma_1$.
The Euclidean chain length $K{+}1$ with terminal remainder~$1$ implies
$\sigma_1\ge F_{1,K+1}$. Otherwise, the chain would be shorter than that
generated by the consecutive Fibonacci pair, contradicting the well-known
minimality of Fibonacci numbers for Euclidean length. Hence,
\begin{equation}\label{eq:gap-lb}
\Gamma_2\ \ge\ \Gamma_1+F_{1,K+1}\ \ge\ \lceil\rho\rceil+F_{1,K+1}.
\end{equation}

\medskip
\noindent\emph{Case $K=1$.}
Here $F_{1,2}=2$, so \eqref{eq:gap-lb} gives $\Gamma_2\ge\Gamma_1+2$.
\begin{itemize}
\item If $\lceil\rho\rceil$ is odd, take $\Gamma_1=\lceil\rho\rceil$ and
      $\Gamma_2=\Gamma_1+2$; then $\gcd(\Gamma_1,\Gamma_2)=1$ and the bound is tight.
\item If $\lceil\rho\rceil$ is even, $\Gamma_1$ must be odd to ensure coprimality
      with $\Gamma_2=\Gamma_1+2$; choose $\Gamma_1=\lceil\rho\rceil+1$ and
      $\Gamma_2=\Gamma_1+2=\lceil\rho\rceil+3$. Reducing $\Gamma_2$ to
      $\lceil\rho\rceil+2$ forces $\Gamma_1=\lceil\rho\rceil$ (even) and breaks
      coprimality.
\end{itemize}
Thus the optimal pair is $\Gamma_1=2\zeta_1+1$ and $\Gamma_2=\Gamma_1+2$, with
$\zeta_1=\lceil(\rho-1)/2\rceil$.

\medskip
\noindent\emph{Case $K=2$.}
Now $F_{1,3}=3$, hence
\begin{equation}\label{eq:K2-gap}
\Gamma_2\ \ge\ \Gamma_1+3\ \ge\ \lceil\rho\rceil+3.
\end{equation}
We consider $\lceil\rho\rceil\equiv r\pmod{3}$.

\smallskip
\underline{$r=2$}: Take $\Gamma_1=\lceil\rho\rceil$ and
set $\Gamma_2=\Gamma_1+3$; \eqref{eq:K2-gap} is met with equality and the
remainder chain is $(3,2,1)$.

\smallskip
\underline{$r=1$}: If we tried $\Gamma_2=\lceil\rho\rceil+3$ with
$\Gamma_1=\lceil\rho\rceil$, then $\sigma_1=3$ and the remainder chain would be $(\sigma_1,\sigma_2)=(3,1)$,
i.e., only $K=1$. Thus we must increase $\Gamma_1$ to keep $K=2$ while
respecting coprimality. Taking $\Gamma_1=\lceil\rho\rceil+1$ and
$\Gamma_2=\Gamma_1+3=\lceil\rho\rceil+4$ yields the chain $(3,2,1)$ and $\Gamma_2$ is
minimal.

\smallskip

\underline{$r=0$:} Write $\lceil\rho\rceil=3k$. If $k\equiv1\pmod{4}$, then
$\lceil\rho\rceil\equiv3\pmod{12}$, i.e., $\lceil\rho\rceil=12\ell+3$ with $k=4\ell+1$.
In this subcase, the choice $(\Gamma_1,\Gamma_2)=(12\ell+3,\,12\ell+7)$ is optimal as
reducing $\Gamma_2$ to $12\ell+6$ would give $\gcd(\Gamma_1,\Gamma_2)=3$, breaking
coprimality.

For the remaining classes $\lceil\rho\rceil\in\{12\ell,\,12\ell+6,\,12\ell+9\}$, take
\[
\Gamma_1=\lceil\rho\rceil+2,\qquad
\Gamma_2=\Gamma_1+3=\lceil\rho\rceil+5,
\]
which yields the remainder chain $(\sigma_1,\sigma_2,\sigma_3)=(3,2,1)$ (hence $K=2$).
Any smaller second modulus fails:
(\emph{i}) $\Gamma_2=\lceil\rho\rceil+3$ forces $\Gamma_1=\lceil\rho\rceil$, so
$\gcd(\Gamma_1,\Gamma_2)=3$;
(\emph{ii}) $(\Gamma_1,\Gamma_2)=(\lceil\rho\rceil+1,\lceil\rho\rceil+4)$ collapses the chain to $(3,1)$ (only $K=1$);
(\emph{iii}) $(\Gamma_1,\Gamma_2)=(\lceil\rho\rceil,\lceil\rho\rceil+4)$ gives $(4,1)$ when
$\lceil\rho\rceil=12\ell+9$ ($K=1$), and violates coprimality when
$\lceil\rho\rceil\in\{12\ell,\,12\ell+6\}$ since $\lceil\rho\rceil$ and $\lceil\rho\rceil+4$ are both even (i.e., they have a common divisor of 2).
Hence $\Gamma_2=\lceil\rho\rceil+5$ is minimal in these subcases.
\end{IEEEproof}

\bibliographystyle{IEEEtran}
\bibliography{refs}

\vfill

\end{document}